\documentclass[fleqn,usenatbib]{mnras}

\usepackage{newtxtext,newtxmath}

\usepackage[T1]{fontenc}

\DeclareRobustCommand{\VAN}[3]{#2}
\let\VANthebibliography\thebibliography
\def\thebibliography{\DeclareRobustCommand{\VAN}[3]{##3}\VANthebibliography}

\usepackage{graphicx}	
\usepackage{amsmath}	
\usepackage{orcidlink}
\usepackage{tablefootnote}
\usepackage{caption}
\usepackage{upgreek}
\usepackage{multicol}
\usepackage{multirow}

\DeclareCaptionLabelFormat{notes}
{%
    #1 Notes}

\defcitealias{Narayanan_2014}{NK14}

\title[CO excitation in high-$z$ SMGs]{The properties of the interstellar medium in dusty, star-forming galaxies at ${\emph z} \sim 2-4$: The shape of the CO spectral line energy distributions}

\author[D. J. Taylor et al.]{Dominic J. Taylor$^{\orcidlink{https://orcid.org/0000-0002-0031-630X}}$,$^{1}$\thanks{E-mail: \href{mailto:dominic.j.taylor@durham.ac.uk}{dominic.j.taylor@durham.ac.uk}} A. M. Swinbank$^{\orcidlink{https://orcid.org/0000-0003-1192-5837}}$,$^1$ Ian Smail$^{\orcidlink{https://orcid.org/0000-0003-3037-257X}}$,$^1$ Annagrazia Puglisi$^{\orcidlink{https://orcid.org/0000-0001-9369-1805}}$,$^2$ Jack E. Birkin$^{\orcidlink{https://orcid.org/0000-0002-3272-7568}}$,$^{3,4}$
\newauthor Ugne Dudzevi{\v c}i{\= u}t{\. e}$^{\orcidlink{https://orcid.org/0000-0003-4748-0681}}$,$^5$ Chian-Chou Chen$^{\orcidlink{https://orcid.org/0000-0002-3805-0789}}$,$^6$ S. Ikarashi,$^{7,8}$ Marta Frias Castillo,$^{9}$ Axel Wei{\ss}$^{\orcidlink{https://orcid.org/0000-0003-4678-3939}}$,$^{10}$
\newauthor Zefeng Li$^{\orcidlink{https://orcid.org/0000-0001-7373-3115}}$,$^1$ Scott C. Chapman,$^{11}$ Jasper Jansen,$^{9}$ E. F. Jim{\' e}nez-Andrade$^{\orcidlink{https://orcid.org/0000-0002-2640-5917}}$,$^{12,13,14}$ Leah K. Morabito$^{\orcidlink{https://orcid.org/0000-0003-0487-6651}}$,$^{1,15}$
\newauthor Eric J. Murphy$^{\orcidlink{https://orcid.org/0000-0001-7089-7325}}$,$^{12}$ Matus Rybak$^{\orcidlink{https://orcid.org/0000-0002-1383-0746}}$,$^{9,16,17}$ and P. P. van der Werf$^{\orcidlink{https://orcid.org/0000-0001-5434-5942}}$$^{9}$
\\
$^{1}$Centre for Extragalactic Astronomy, Department of Physics, Durham University, South Road, Durham DH1 3LE, UK \\
$^2$School of Physics and Astronomy, University of Southampton, Highfield SO17 1BJ, UK \\
$^3$Department of Physics and Astronomy, Texas A\&M University, 4242 TAMU, College Station, TX 77843-4242, USA \\
$^4$George P. and Cynthia Woods Mitchell Institute for Fundamental Physics and Astronomy, Texas A\&M University, 4242 TAMU, College Station, \\
TX 77843-4242, USA \\
$^5$Max-Planck-Institut f{\" u}r Astronomie, K{\" o}nigstuhl 17, D-69117 Heidelberg, Germany \\
$^6$Academia Sinica Institute of Astronomy and Astrophysics (ASIAA), No. 1, Sec. 4, Roosevelt Road, Taipei 10617, Taiwan \\
$^7$Department of Physics, General Studies, College of Engineering, Nihon University, 1 Nakagawara, Tokusada, Tamuramachi, Koriyama, Fukushima, \\
963-8642, Japan \\
$^8$National Astronomical Observatory of Japan, 2-21-1 Osawa, Mitaka, Tokyo, 181-8588, Japan \\
$^{9}$Leiden Observatory, Leiden University, P.O. Box 9513, 2300 RA Leiden, The Netherlands \\
$^{10}$Max Planck Institut f{\" u}r Radioastronomie, Auf dem H{\" u}gel 69, D-53121 Bonn, Germany \\
$^{11}$Department of Physics and Atmospheric Science, Dalhousie University, Halifax, Halifax, NS B3H 3J5, Canada \\
$^{12}$National Radio Astronomy Observatory, 520 Edgemont Road, Charlottesville, VA 22903, USA \\
$^{13}$Instituto de Radioastronom\'{i}a y Astrof\'{i}sica, Universidad Nacional Aut{\' o}nomia de M\'{e}xico, Antigua Carretera a P\'{a}tzcuaro \# 8701, Ex-Hda. San Jos\'{e} de la \\
Huerta, Morelia, Michoac\'{a}n, C.P 58089, M\'{e}xico \\
$^{14}$Argelander Institut f\"{u}r Astronomie, Universit\"{a}t Bonn, Auf dem H\"{u}gel 71, Bonn, D-53121, Germany \\
$^{15}$Institute for Computational Cosmology, Department of Physics, University of Durham, South Road, Durham DH1 3LE, UK \\
$^{16}$Faculty of Electrical Engineering, Mathematics and Computer Science, Delft University of Technology, Mekelweg 4, 2628 CD Delft, the Netherlands \\
$^{17}$SRON - Netherlands Institute for Space Research, Niels Bohrweg 4, 2333 CA Leiden, The Netherlands
}

\date{Accepted XXX. Received YYY; in original form ZZZ}

\pubyear{2024}

\begin{document}
\label{firstpage}
\pagerange{\pageref{firstpage}--\pageref{lastpage}}
\maketitle

\begin{abstract}
The molecular gas in the interstellar medium (ISM) of star-forming galaxy populations exhibits diverse physical properties.
We investigate the $^{12}$CO excitation of twelve dusty, luminous star-forming galaxies at $z \sim$\,2--4 by combining observations of the $^{12}$CO from $J_{\rm up} = 1$ to $J_{\rm up} = 8$.
The spectral line energy distribution (SLED) has a similar shape to NGC 253, M82, and local ULIRGs, with much stronger excitation than the Milky Way inner disc.
By combining with resolved dust continuum sizes from high-resolution $870$-$\upmu$m ALMA observations and dust mass measurements determined from multi-wavelength SED fitting, we measure the relationship between the $^{12}$CO SLED and probable physical drivers of excitation: star-formation efficiency, the average intensity of the radiation field $\langle U\rangle$, and the star-formation rate surface density. The primary driver of high-$J_{\rm up}$ $^{12}$CO excitation in star-forming galaxies is star-formation rate surface density.
We use the ratio of the CO(3--2) and CO(6--5) line fluxes to infer the CO excitation in each source
and find that the average ratios for our sample are elevated compared to observations of low-redshift, less actively star-forming galaxies and agree well with predictions from numerical models that relate the ISM excitation to the star-formation rate surface density.
The significant scatter in the line ratios of a factor $\approx 3$ within our sample likely reflects intrinsic variations in the ISM properties which may be caused by other effects on the excitation of the molecular gas, such as cosmic ray ionization rates and mechanical heating through turbulence dissipation.
\end{abstract}




\section{Introduction}


In the local Universe, Ultra-Luminous Infrared Galaxies (\citealp[ULIRGs;][]{Neugebauer_1984}) represent a population of highly-obscured, dusty star-forming galaxies (DSFGs) with high far-infrared luminosities ($L_{\rm IR} \geq 10^{12}$\,L$_\odot$) and inferred star-formation rates (SFR) of ${\rm SFR} \geq 100$\,M$_\odot$yr$^{-1}$ resulting from the compression and cooling of gas triggered by major mergers \citep{Sanders_1996}.  Although they contribute only a small fraction of the total star-formation rate density at $z \sim 0$, the contribution of ULIRGs at $z \gtrsim 1$ is significantly higher
\citep[e.g.,][]{Murphy_2011,Magnelli_2013,Dudzeviciute_2020} at least out to $z \sim 5$ \citep[e.g.,][]{Bouwens_2020}.

Submillimetre galaxies \citep[SMGs;][]{Smail_1997,Barger_1998,Hughes_1998,Eales_1999} represent a subset of
high-redshift ($z \approx $\,1--5) DSFGs that are over two orders of magnitude more numerous than local ULIRGs \citep[e.g.,][]{Dudzeviciute_2020}, with star-formation rates (${\rm SFR} \gtrsim 100$ M$_\odot$yr$^{-1}$) and dust masses ($M_{\rm dust} \gtrsim 10^8$ M$_\odot$) similar to those of the most luminous ULIRGs at $z\sim$\,0.
The submillimetre emission arises primarily from the continuum emission from the reprocessing by obscuring dust grains of the ultraviolet (UV) light emitted by young stars, with line emission from atomic and molecular transitions in the interstellar gas superimposed (see \citealt{Casey_2014a} for a review).
Their high star-formation rates and strong dust obscuration results in the majority of their optical/UV light being absorbed and re-emitted in the infrared, producing far-infrared luminosities of $L_{\rm IR} \gtrsim 10^{12-13}$~L$_\odot$.
Moreover, submillimetre emission from SMGs appears enhanced at higher redshifts by virtue of a negative $K$-correction.
Within the interstellar medium (ISM) of SMGs, star formation appears to occur within compact regions ($\sim 2-3$\,kpc diameter) that are highly embedded in dust \citep[e.g.,][]{Tacconi_2006,Younger_2010,Simpson_2015,Ikarashi_2015,Hodge_2016,Gullberg_2019}.
The compact star formation may be the result of mergers or interactions, which have driven the gas to the centre \citep[e.g.,][]{Tacconi_2006} or due to instabilities driven by high gas fractions or external torques \citep[e.g.,][]{Hodge_2016,Hodge_2019,Gullberg_2019}.
Their intensely star-forming nature, and emission in the submillimetre regime
makes SMGs ideal laboratories for investigating the process of star formation at high redshifts \citep[e.g.,][]{Danielson_2011,Danielson_2013}.


The bulk of the gas inside the giant molecular clouds (GMCs) from which stars form is comprised of molecular hydrogen, H$_2$. However, this dominant molecule is difficult to observe due to its lack of a permanent dipole moment. This means that at low temperatures H$_2$ does not have detectable rotational transitions; the first quadrupole-moment lines require temperatures of $\sim 500$~K to be excited. However, GMCs in typical star-forming galaxies have temperatures of $\sim 100$~K near star-forming regions inside photodissociation regions (PDRs) \citep{Fukui_2010,Krumholz_2014,Heyer_2015}. 
Therefore, any detected emission through higher-order transitions of H$_2$ is not directly tracing the physical conditions of the clouds in which stars are likely to be forming. Regardless, even for shocked or strongly irradiated H$_2$ gas in PDRs where ground state excitations can occur, there is strong atmospheric absorption at its rest wavelength of $\sim 28\upmu$m making it unobservable with current facilities.
Moreover, $\Delta J = J_{\rm up} - J_{\rm lo} = 2$ rotational H$_2$ lines also provide important information but only for a very small fraction of the mass of a GMC ($\simeq 1\%$).
Hence, the second most-abundant molecular species, carbon monoxide ($^{12}$CO, hereafter CO), is commonly used instead as a tracer of the H$_2$ reservoirs since it produces detectable line emission from rotational transitions, allowing us to
map the range of temperatures and densities present in the ISM.
Excitation to higher states can be caused by a combination of collisions with H$_2$ and He, as well as through radiative absorption. 
The lowest $J = $\,1--0 transition requires a minimum temperature $T = 5.5$\,K for significant excitation, while higher-$J_{\rm up}$ transitions trace warmer gas, such as the $J = 6$--5 transition which requires $T \sim 120$\,K.
At a fixed kinetic temperature $T = 40$\,K and ortho-H$_2$:para-H$_2$ ratio of 3, the $J = $\,1--0 transition in an optically thin gas corresponds to a critical density of $n_{\rm crit} \sim 3 \times 10^2$\,cm$^{-3}$, while the $J = $\,6--5 transition corresponds to $n_{\rm crit} \sim 8 \times 10^4$\,cm$^{-3}$ \citep{Greve_2014}.
CO transitions of $J = $3--2 and lower are usually excited in both star-forming and quiescent H$_2$ gas, due to their sufficient critical densities and temperatures. Thus, higher-$J$ CO lines must be observed for the star-forming H$_2$ gas phase to be selected.

The CO line luminosities for different $J_{\rm up}$-rotational transitions (commonly known as the Spectral Line Energy Distribution, SLED) can yield constraints on the average H$_2$ gas density and temperature of the bulk of the ISM in galaxies.
The relative amounts of warm/dense gas with respect to the cold/diffuse gas (lower rotational transition numbers) can be indicators of the fraction of gas associated with star formation activity, the average conditions prevailing in various H$_2$ gas phases, and the power sources maintaining them.
The CO SLED up to $J_{\rm up} = 3$ provides little constraint on these issues because of the strong density-temperature SLED degeneracies.
Observations of higher CO transitions ($J_{\rm up} \geq 3$) are therefore critical in order to reduce these CO SLED uncertainties and better assess the star-forming dense/warm H$_2$ mass, and its average conditions.
Here we must note that the aforementioned density-temperature degeneracies also affect the high-$J$ CO lines. However, typically for CO lines higher than $J = $\,3--2, the global CO(high-$J$)/(low-$J$) ratios are much more sensitive to the (warm-dense star-forming H$_2$ mass)/(total-H$_2$ mass) gas mass fractions.

\renewcommand{\arraystretch}{1.25}
\begin{table*}
\centering
    \caption{Summary of the targets in our sample of SMGs including the target ALMA ID, R.A, Dec., spectroscopic redshift ($z_{\rm spec}$), dust continuum $870$-$\upmu$m flux density ($S_{870}$), infrared luminosity ($L_{{\rm FIR}[8-1000\upmu {\rm m}]}$), stellar mass ($M_{\ast}$), dust temperature ($T_{\rm d}$), star-formation rate (SFR), effective radius ($R_{\rm eff}$), and star-formation rate surface density ($\Sigma_{\rm SFR}$).}
    \begin{tabular}{@{}cccccccccccccc@{}}
    \hline\hline
    Target & R.A & Dec. & ${z_{\rm spec}}^a$ & $S_{870}$ & ${L_{\rm IR}}^b$ & log$_{10}$(M$_{\ast})^b$ & ${T_{\rm d}}^b$ & SFR$^b$ & ${R_{\rm eff}}^c$ & $\Sigma_{\rm SFR}$ \\
     & (J2000) & (J2000) &  & (mJy) & ($10^{12} L_\odot$) & (M$_\odot$) & (K) & (M$_\odot$ yr$^{-1}$) & (kpc) & (M$_\odot$ yr$^{-1}$ kpc$^{-2}$) \\
    \hline
    AS2UDS009.0      & 02:16:43.77 & $-$05:17:54.7 & 2.942 & $10.1 \pm 0.6$ & $6.6 \pm 2.0$  & $11.4 \pm 0.1$ & $32 \pm 6$ & $700_{-10}^{+170}$   & $1.2 \pm 0.1$ & $79_{-5}^{+20}$     \\
    AS2UDS011.0      & 02:16:30.77 & $-$05:24:02.6 & 4.074 & $11.1 \pm 0.7$ & $8.9 \pm 5.2$  & $11.5 \pm 0.2$ & $43 \pm 8$ & $960_{-190}^{+170}$  & $0.7 \pm 0.1$ & $310_{-70}^{+60}$   \\
    AS2UDS012.0$^d$  & 02:18:03.57 & $-$04:55:26.9 & 2.521 & $10.3 \pm 0.7$ & $3.8 \pm 1.2$  & $11.3 \pm 0.2$ & $31 \pm 2$ & $400_{-90}^{+90}$    & $1.3 \pm 0.2$ & $37_{-10}^{+10}$    \\
    AS2UDS014.0      & 02:17:44.29 & $-$05:20:08.9 & 3.805 & $11.9 \pm 0.6$ & $7.9 \pm 1.6$  & $11.1 \pm 0.1$ & $36 \pm 6$ & $690_{-40}^{+160}$   & $0.7 \pm 0.1$ & $210_{-20}^{+50}$   \\
    AS2UDS026.0      & 02:19:02.24 & $-$05:28:56.6 & 3.296 & $10.0 \pm 0.7$ & $5.9 \pm 2.3$  & $11.5 \pm 0.2$ & $41 \pm 7$ & $350_{-100}^{+80}$   & $1.1 \pm 0.1$ & $42_{-13}^{+10}$    \\
    AS2UDS072.0$^d$  & 02:18:37.05 & $-$05:17:54.8 & 3.542 & $8.2  \pm 0.8$ & $5.9 \pm 1.8$  & $10.2 \pm 0.1$ & $35 \pm 6$ & $130_{-30}^{+30}$    & $0.5 \pm 0.1$ & $78_{-21}^{+21}$    \\
    AS2UDS126.0      & 02:15:46.99 & $-$05:18:52.2 & 2.436 & $11.2 \pm 0.4$ & $9.5 \pm 6.1$  & $11.8 \pm 0.4$ & $38 \pm 8$ & $690_{-260}^{+340}$  & $0.7 \pm 0.2$ & $200_{-90}^{+110}$  \\
    AS2COS0009.1$^d$ & 10:00:28.72 & $+$02:32:03.6 & 2.259 & $13.1 \pm 0.3$ & $13.2 \pm 0.8$ & $10.3 \pm 0.2$ & $39 \pm 3$ & $260_{-30}^{+30}$    & $1.1 \pm 0.1$ & $32_{-4}^{+4}$      \\
    AS2COS0014.1$^d$ & 10:01:41.04 & $+$02:04:04.9 & 2.921 & $14.8 \pm 0.4$ & $4.2 \pm 0.4$  & $11.4 \pm 0.1$ & $34 \pm 4$ & $1000_{-90}^{+140}$  & $0.9 \pm 0.1$ & $210_{-40}^{+50}$   \\
    AS2COS0044.1     & 09:59:10.33 & $+$02:48:55.7 & 2.579 & $13.5 \pm 0.3$ & $6.6 \pm 2.0$  & $11.1 \pm 0.2$ & $32 \pm 6$ & $210_{-30}^{+20}$    & $0.9 \pm 0.1$ & $40_{-9}^{+8}$    \\
    AS2COS0065.1     & 09:58:40.29 & $+$02:05:14.7 & 2.414 & $12.6 \pm 0.3$ & $9.1 \pm 1.1$  & $11.3 \pm 0.2$ & $37 \pm 4$ & $490_{-10}^{+10}$    & $0.7 \pm 0.1$ & $170_{-40}^{+40}$   \\
    AS2COS0139.1     & 09:58:19.79 & $+$02:36:10.1 & 3.292 & $12.8 \pm 0.3$ & $8.9 \pm 2.8$  & $11.1 \pm 0.2$ & $43 \pm 8$ & $1030_{-100}^{+100}$ & $0.8 \pm 0.2$ & $290_{-70}^{+70}$ \\
    \hline\hline
    \end{tabular}
    \captionsetup{labelformat=notes}
    \caption*{
    
    $^a$Measured spectroscopic redshifts from our Gaussian fitting. Values for some sources have changed compared to the measurements from \citet{Birkin_2021} and \citet{Liao_2024} but are within $1\sigma$.
    
    $^b$Measured from {\sc magphys} using spectroscopic redshifts. 
    
    $^c$Dust ($870$-$\upmu$m) continuum size from \citet{Gullberg_2019} or Ikarashi et al. (2024, in prep.).
    
    $^d$Candidate hosts of AGN (see Section~\ref{sec:sample}) on the bases of X-ray counterpart detection, SED fitting with an AGN component \citep{Liao_2024}, or inspection of their multi-wavelength SEDs \citep{Dudzeviciute_2020}.}
    \captionsetup{labelformat=default}
    \label{tab:summary}
\end{table*}

Several efforts have been made to parameterise the CO SLEDs of star-forming galaxies as a means to predict lower-order transitions from higher-order, more easily observed transitions.
In the local context, \citet{Papadopoulos_2012} studied variations in the CO SLEDs of $z \leq 0.1$ luminous infrared galaxies (LIRGs)
and suggested that the properties of the high-$J_{\rm up}$ CO lines are due to gas heated by cosmic rays and/or turbulence.
The evolution of CO excitation up to and above $J_{\rm up} = 5$ has been predicted and observationally tested in $z = $\,1.5--2 main sequence star-forming disc galaxies using high-resolution numerical hydrodynamic simulations and observations \citep{Bournaud_2015,Daddi_2015}, attributing high-$J_{\rm up}$ excitation to giant molecular clumps in the disc. Enhanced CO excitation at $z \approx 2$ has also been predicted using semi-analytic models \citep{Lagos_2012,Popping_2014}.

Combining numerical simulations of disc galaxies and mergers with molecular line radiative transfer calculations to acquire level populations of CO emission lines up to $J_{\rm up} = 9$, \citet[][hereafter NK14]{Narayanan_2014} developed a general model to describe the shape of the CO SLED in star-forming galaxies.
In their model, the shape of the SLED correlates well with both the resolved and galaxy-averaged star-formation rate surface density ($\Sigma_{\rm SFR}$), i.e. galaxies with higher $\Sigma_{\rm SFR}$ have more highly excited CO SLEDs.
Empirical conclusions have also been made using CO observations up to $J_{\rm up} = 5$ of $z \sim $\,1.0--1.7 main sequence and starburst galaxies \citep{Daddi_2015,Valentino_2020}.
Resolved observations of a $z = $\,2.3 strongly lensed galaxy, on the other hand, have tentatively suggested that there is no correlation between CO excitation and $\Sigma_{\rm SFR}$ in this galaxy \citep{Sharon_2019}, though only up to a $J_{\rm up} = 3$ and over a narrow range of $\Sigma_{\rm SFR}$.
\citet{Daddi_2015}, however, found a closer correlation between CO excitation and the average intensity of the radiation field $\langle U \rangle$, which is reflected in dust temperature \citep{Chanial_2007,Hwang_2010,Elbaz_2011}.
Other works have instead found that CO excitation in $z = $\,1.0--1.7 main-sequence and starburst galaxies \citep{Valentino_2020}, and $z \sim $\,2--3 SMGs \citep{Sharon_2016}, correlates strongest with star-formation efficiency (SFE = SFR / M$_{\rm gas}$).

Studies of the dependence of the shape of CO SLEDs so far have been limited in the precision and range of measured $\Sigma_{\rm SFR}$ for their samples and to low-$J_{\rm up}$ CO transitions used as a tracer of excitation in SMGs at high redshifts. They have also typically been carried out on lower redshift ($z \leq 2$) galaxies.
This study aims to expand the parameter space explored in the investigation of the correlation between SLED shape and the possible physical drivers, with a particular focus on $\Sigma_{\rm SFR}$, to higher $J_{\rm up}$ and higher redshifts.

In this paper, we exploit new and existing Atacama Large Millimeter/submillimeter Array (ALMA) observations of the mid- to high-$J_{\rm up}$ CO transitions ($J_{\rm up}$\,=\,3 or 4 and $J_{\rm up}$\,=\,6 or 8) from twelve SMGs at $z$\,=\,2--4, along with existing $J_{\rm up}$\,=\,1 measurements from the Karl G. Jansky Very Large Array (JVLA) to study the excitation of the ISM, testing the various model predictions which attempt to describe variations in the CO SLED, in particular those from \citetalias{Narayanan_2014}.
In Section~\ref{sec:sample}, we describe the sample selection and reduction of the new CO(6--5) or CO(8--7) observations. In Section~\ref{sec:analysis}, we present the resulting spectra and compare them to other well-studied systems, both local and at high redshifts, and determine the correlations between CO SLED properties and physical drivers. In Section~\ref{sec:discussion}, we discuss the relation between CO SLED shape and the possible physical drivers, particularly star-formation rate surface density, comparing our findings with theoretical models, and propose possible explanations for the differences found between observations and theory.
In Section~\ref{sec:conclusions}, we summarise our conclusions.

We adopt a flat $\Lambda$-CDM cosmology with $\Omega_{\rm m} = 0.32$, $\Omega_\Lambda = 0.68$, and H$_0 = 67$ km s$^{-1}$ Mpc$^{-1}$ \citep{Planck_2018}.


\section{Sample, Observations and Data Reduction}
\label{sec:sample}

\subsection{Sample Selection}

To study the excitation within the ISM of star forming galaxies, we selected twelve spectroscopically-confirmed, unlensed SMGs at $z$\,=\,2.3--4.1 from the AS2UDS \citep{Stach_2019} and AS2COSMOS \citep{Simpson_2020} surveys.
AS2UDS is an ALMA $870\upmu$m study of $\approx 700$ sources selected from the SCUBA-2 Cosmology Legacy Survey \citep[S2CLS][]{Geach_2017} map of the UKIRT Infrared Deep Sky Survey (UKIDSS) UDS, while AS2COSMOS used ALMA at $870\upmu$m to follow up 180 of the brightest 850$\upmu$m sources from the SCUBA-2 map of the COSMOS field from \citet{Simpson_2019}.
All of the galaxies in our sample have existing measurements of the CO(3--2) or CO(4--3) emission line luminosities \citep{Birkin_2021,Chen_2022}, and all have CO(1--0) measurements (or robust upper limits) from JVLA (\citealt{FriasCastillo_2023}, Jansen et al. 2024, in prep.).
The SMGs in this sample are by selection some of the brightest $870$-$\upmu$m sources in the parent samples, with a median $870$-$\upmu$m flux density ($S_{870\upmu{\rm m}}$) of 11.5~mJy and a range of $S_{870\upmu{\rm m}} \simeq $\,8--15~mJy. 
The sample galaxies have cold dust masses in the range $10^{9.0-9.7}$ M$_\odot$, with a median of $10^{9.3}$ M$_\odot$, and infrared luminosities\footnote{Measured across the range $\uplambda = $\,8--1000$\upmu$m.} in the range $L_{\rm IR} \sim 10^{12.5-13.4}$~L$_\odot$, with a median of $10^{12.9}$~L$_\odot$.
A summary of the properties of the sample can be found in Table~\ref{tab:summary}.

To calculate the star-formation rate surface densities, we adopt the star-formation rates derived from a {\sc magphys} \citep{daCunha_2008,daCunha_2015}
analysis of the source spectral energy distributions (SEDs) using the CO-derived redshift and the multi-wavelength photometry, including the Herschel/SPIRE 250--500$\upmu$m and ALMA $870\upmu$m fluxes in the far-infrared to constrain the far-infrared SEDs. Star-formation rates of AS2UDS sources are presented in \citet{Birkin_2021} with star-formation rates of sources in AS2COSMOS reported in \citet{Liao_2024}.
For our sample, the median and quartile ranges are ${\rm SFR} = 590 \pm 440$\,M$_\odot$yr$^{-1}$.
The spatial extent of the dust continuum emission was measured for AS2UDS sources from
ALMA observations by \citet{Gullberg_2019} and for AS2COSMOS sources
by Ikarashi et al. (2024, in prep.).
These observations constrain the spatial extent of the dust continuum, with a range of sizes (half light radii) of 0.5--1.3\,kpc, with a median of 0.7\,kpc (Table~\ref{tab:summary}).
To derive the star-formation rate surface density, we adopt the half light radii of the dust (and correspondingly half of the total star-formation rate).

The sample spans a range in $\Sigma_{\rm SFR}$ of an order of magnitude with a median and 16--84$^{\rm th}$ percentile range of $\Sigma_{\rm SFR} \sim 140$\,(60--330) M$_\odot$~yr$^{-1}$~kpc$^{-2}$,
while having similar dust temperatures with a median $T_{\rm d} = 37 \pm 4$ K.
Compared to the parent AS2UDS and AS2COSMOS samples, our subsample has bright dust continuum, and a typical $\Sigma_{\rm SFR}$ which is a factor of 1.5 times the average of $\Sigma_{\rm SFR} \sim 80$\,M$_\odot$~yr$^{-1}$~kpc$^{-2}$ and similar dust temperatures to the average of $T_{\rm d} = 40$\,K \citep{Simpson_2020,Stach_2019}.

\begin{figure}
\centering
	\includegraphics[width=1\columnwidth]{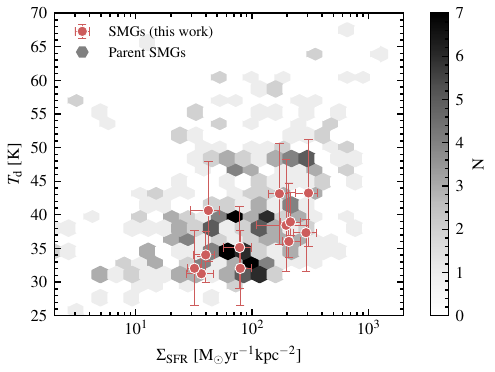}
    \caption{The distribution of dust temperatures ($T_{\rm d}$) and star-formation rate surface densities ($\Sigma_{\rm SFR}$) for the SMGs in our ALMA sample, along with the parent sample \citep{Dudzeviciute_2020}, given by 2-D hexbins with number densities indicated by the colour bar. Though the SMGs in our sample represent the subset of the SMG population that are some of the brightest in their dust continuum, their dust temperatures and $\Sigma_{\rm SFR}$ are representative of the parent population.}
    \label{fig:Td_SigmaSFR}
\end{figure}

The SMGs in our sample were selected on the basis of their high dust luminosities but are, nonetheless, representative of the parent SMG population with respect to both their dust temperatures and $\Sigma_{\rm SFR}$. We highlight this in Figure~\ref{fig:Td_SigmaSFR}, where we show the distribution for the sample studied in this work against the parent SMG sample.

\subsection{AGN contamination}
\label{sec:agn}

It is important to identify whether any of the
line ratios and properties of the physical drivers of sources in our sample could include a contribution of excitation from AGN processes.
The CO SLEDs of sources with bolometric luminosities that are contributed-to most significantly by star formation, and with normal X-ray luminosities \citep[$L_X \leq 10^{-4} \times L_{\rm IR}$;][]{Alexander_2005}, are not significantly affected up to mid-$J_{\rm up}$.
In addition, existing SLED models \citep{Narayanan_2014} are able to already replicate the excitation ladders of nearby active galaxies such as Mrk 231 \citep{van_der_Werf_2010} and NGC 6240 \citep{Meijerink_2013} up to $J_{\rm up} = 9$, suggesting that the heating provided by star-formation is sufficient to explain the emission up to high-$J_{\rm up}$ without the inclusion of AGN-driven heating, although CO excitation is predicted by semi-analytic models to increase at $J_{\rm up} > 6$ \citep{Lagos_2012}.
Moreover, for X-ray dominated regions to encompass the bulk of the H$_2$ gas reservoir of a galaxy, it must be very closely located around the AGN.
In the X-ray luminous AGN Mrk 231 ($L_{X}[$2--10 keV$] \sim 5 \times 10^{43}$ erg/s), for example, the X-ray dominated H$_2$ gas reservoir is located within an $r \sim 160$\,pc radius \citep{van_der_Werf_2010}.
CO SLEDs of quasi-stellar objects (QSOs), on the other hand, that have bolometric luminosities dominated by the AGN, are expected to peak at higher-$J_{\rm up}$ \citep{Pensabene_2021} and may have relatively extended dust continuum sizes \citep{Ikarashi_2017}, which are dependent on their surface brightness profiles, resulting in apparently increased star-formation rate surface densities.

Of the sample of twelve SMGs, one (AS2COS0014.1) has an X-ray counterpart with $L_X \sim 10^{43}$\,erg s$^{-1}$ and another (AS2COS009.1) has been identified from multi-wavelength SED fitting with the consideration of an AGN component as having $> 5\%$ contribution to its total luminosity by an AGN \citep{Liao_2024}, while two (AS2UDS012.0 and AS2UDS072.0) have mid-infrared SEDs \citep{Dudzeviciute_2020} with power-law shapes that could indicate the presence of an AGN in the mid-infrared emission. Although none have been identified on the basis that their X-ray luminosities exceed the $L_X = 0.004 \times L_{\rm IR}$ limit which is the typical threshold used to classify SMGs as AGN \citep{Alexander_2005,Chapman_2024}, in the following sections we consider these four sources as candidate hosts of AGN in order to evaluate how any contribution from an AGN may affect the CO SLEDs.

\begin{figure*}
\centering
	\includegraphics[width=1\linewidth]{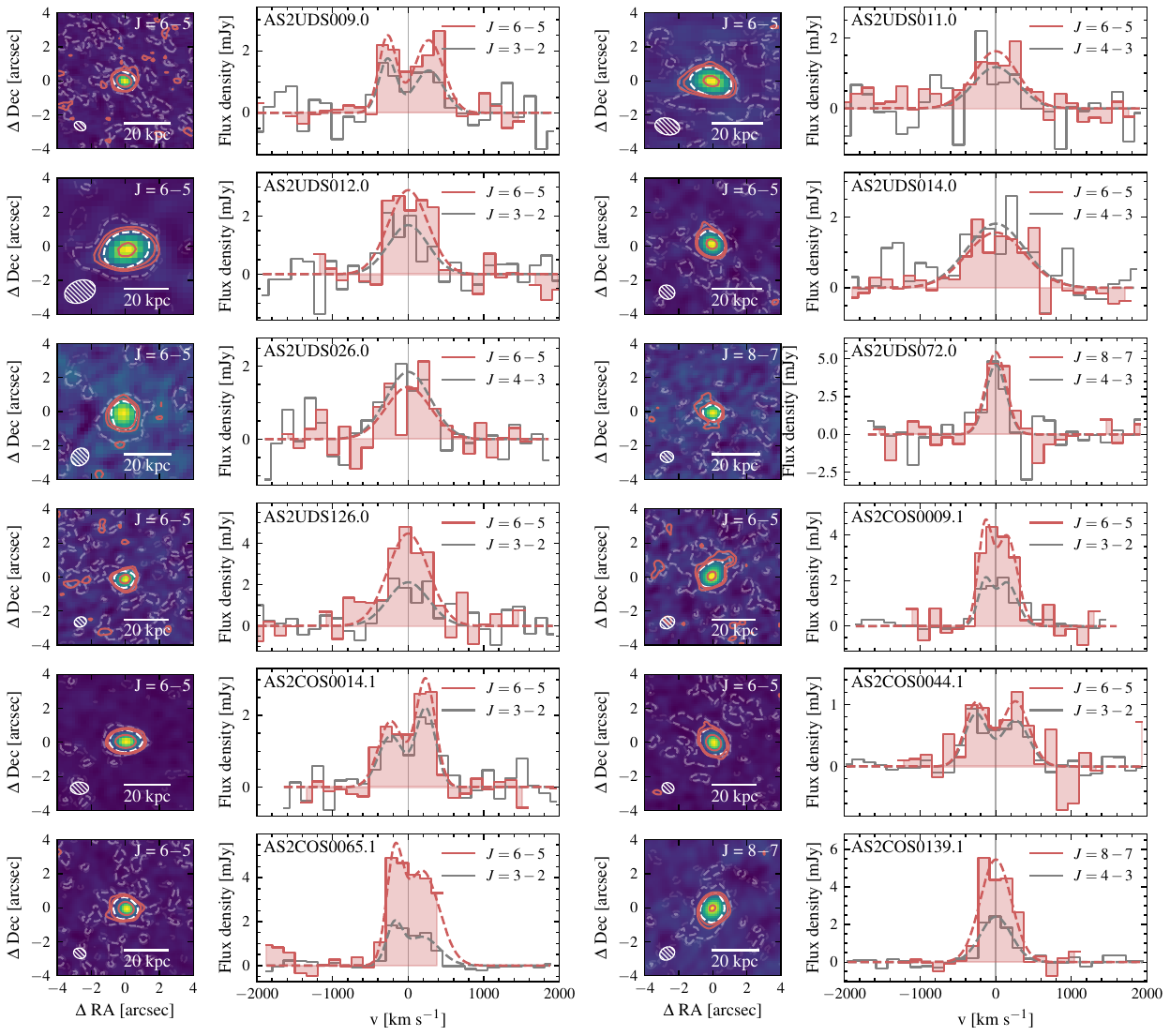}
        \caption{The left-hand of each set of panels shows $4^{\prime\prime} \times 4^{\prime\prime}$ CO(6--5) or CO(8--7) emission line maps with continuum, and 3-, 5-, and 25-$\sigma$ contours of the continuum emission (dashed) and continuum-subtracted CO line emission (solid), with the synthesised beam indicated in the lower-left and scale bar in the lower-right.
        Alongside, are panels showing the emission lines (solid histograms) with Gaussian profile fits (dashed curves) to the continuum-subtracted restframe spectra for the mid-$J_{\rm up}$ and high-$J_{\rm up}$ lines, with channel widths of $150$\,km\,s$^{-1}$.
        As a reference, the systemic redshift $v = 0$\,km\,s$^{-1}$ is indicated by the solid line.
        The CO(6--5) line in AS2COS0065.1 falls at the edge of Band 6, causing it to be cut off.
        The redshift and FWHMs of single Gaussian fits were required to agree between transitions, while for the double Gaussian fits the redshift and FWHM of each component were allowed to vary, but were required to match the same component in the other transition.
        High-$J_{\rm up}$ line fluxes are higher than mid-$J_{\rm up}$ lines, suggesting that the ISMs of SMGs have intense star-forming conditions.
        }
    \label{fig:all_spec_clean_bscaled}
\end{figure*}


\subsection{Observations and Data Reduction}

All of the targets were observed with the ALMA 12-m array in Band 5 or 6 as part of program 2021.1.00666.S, where each observation used a spectral setup of two pairs (two spectral windows) of 3.75\,GHz sidebands to achieve a total bandwidth of 7.5\,GHz. Across the sample, a frequency range of $\sim$ 136--214 GHz was covered. The configuration of the array resulted in a spatial resolution range of 0.7--1.9$^{\prime\prime}$ beam FWHM (Table~\ref{tab:fluxes}).


We calibrated the raw data to obtain the measurement sets from
which the data-cubes were cleaned using the {\sc tclean} algorithm of the Common Astronomy Software Applications package \citep[{\sc casa};][]{CASA_2022} version 6.2.1-7.
To optimise the efficiency of the cleaning, visibilities were transformed to cubes of differing spatial dimensions dependent on the characteristics of the observations: for a given source and SPW, we adopt a spatial sampling (cell) size equal to $1/5$ of the FWHM of the synthesised beam,
and set the \texttt{imsize} based on the cell size, so that the number of cells covers the spatial extent of the full field of view.
We create
a user-defined mask for the cleaning of all the {\it uv}-visibilities in an iterative process, different to the auto-mask approach which would automatically determine a mask shape and size.  We apply a circular mask of radius $1.5$ times the FWHM of the synthesised beam (major-axis diameter) in the pipeline-reduced data-cube from the image centre.
We note that, although the emission from the source may vary from channel-to-channel, the size of the mask is fixed to ensure that all the emission is cleaned and no flux is lost in the process, based on inspection of the curves of growth for the pipeline-reduced products.  We also adopt a natural baseline weighting
resulting in the optimal signal-to-noise (S/N) but the
coarsest resolution.
The threshold that the peak residual within the clean mask across all channels is compared to is set to three times the root-mean-square of the background measured from the cubes pre-cleaning (\texttt{threshold} $= 3\sigma_{\rm RMS, dirty}$). The size of the restoring beam, which is convolved with the model and added to the residuals, was set as the common size appropriate for all channels (\texttt{restoringbeam} = ``common''). Adopting this method of cleaning was able to effectively model and remove (deconvolve) the ``dirty'' beam from the sky image in order to reconstruct the true source brightness, demonstrated by the absence of remnants of the dirty beam in the residual map.

\section{Analysis \& Results}
\label{sec:analysis}

\subsection{Spectral Analysis}

Given the known positions of the sources and spectroscopic redshifts, we create emission and continuum line maps of each source from the cleaned cubes.  To extract the spectra
and measure fluxes,
we adopt apertures centred on the $870$-$\upmu$m ALMA positions of emission determined by \citet{Stach_2019} and \citet{Simpson_2020} for the AS2UDS and AS2COSMOS sources, respectively.

We measure the total flux from the sources for each $J_{\rm up}$ transition so that the resulting CO SLEDs reflect the global properties of the ISM in the galaxies.
To measure the total flux,
we employ an elliptical aperture with the same axis ratio as the synthesised beam, but with three times its size.
The sources are relatively compact; however the maximum recoverable scales of each of the observations using JVLA, NOEMA, or ALMA are sufficiently large enough ($\gtrsim 7$\,arcsec) to contain the most extended emission which is expected.
We calculate and apply aperture corrections to each spectrum using restored images of the calibrators. Given the apertures we adopt, the correction factors range from 1.01--1.16 with a median of 1.06.

For the CO(3--2) or CO(4--3) spectra of AS2COSMOS sources, we use the aperture correction factors calculated by \citet{Chen_2022}, applying a correction of a factor of 1.2 (median of range 1.1--1.3) to the corresponding spectra.
For the CO(3--2) or CO(4--3) spectra of AS2UDS sources, aperture corrections have already been applied \citep{Birkin_2021}.  We show the final spectra in Figure~\ref{fig:all_spec_clean_bscaled} alongside the emission maps used to extract them.
We note that for sources AS2UDS072.0 and AS2COS0139.1, the emission lines in our study are CO(8--7) rather than CO(6--5), due to a misidentification of the original mid-$J_{\rm up}$ CO line.

We measured the continuum using a linear fit to each of the spectra, considering only channels $|v| > 800$\,km~s$^{-1}$, and we report the resulting continuum flux derivatives in Table~\ref{tab:fluxes}.
To measure the emission line fluxes, we fit a Gaussian profile to each of the spectra to be consistent with the method from \citet{FriasCastillo_2023}. For consistency, we use either a single or double Gaussian profile depending on which was used in the analyses by \citet{Birkin_2021} and \citet{Chen_2022}. This was done for all sources, with the exception of AS2COS0139.1, where the fit performed better with a single Gaussian profile and which was subsequently adopted. Therefore, single Gaussian profiles were fitted to the spectra of seven sources in our sample, while double Gaussians were fitted to the spectra of the remaining five sources (see Table~\ref{tab:fluxes}). For the source AS2COS0014.1, which was covered in both studies, we adopt a double Gaussian profile following \citet{Chen_2022}.
To ensure consistency between the low-$J_{\rm up}$ and high-$J_{\rm up}$ CO emission line fluxes, we also remeasure the fluxes of the CO(3--2) and CO(4--3) lines from \citet{Birkin_2021} and \citet{Chen_2022} in the same way. The fluxes we measure agree within 1$\sigma$ of their previously reported values in all cases except for AS2UDS014.0 which agrees within $2\sigma$.

We fit the mid-$J_{\rm up}$ and high-$J_{\rm up}$ lines simultaneously. For single Gaussian profiles, the redshift and FWHM were required to agree between both transitions. For double Gaussian profiles, the redshift and FWHM of each component were allowed to vary, but were required to match the same component in the other transition.
Uncertainties on the measured high-$J_{\rm up}$ FWHMs and fluxes were obtained through a Monte Carlo method, refitting the CO lines after randomly resampling and adding noise from apertures in the background of the maps, a total of 5000 times.
Uncertainties on the measured $J_{\rm up} = $\,3 or 4 FWHMs and fluxes were estimated using the S/N of their associated spectra.
The S/N of the high-$J_{\rm up}$ (CO(6--5) or CO(8--7)) spectra range from 15 to 61 with a median of 26, while the S/N of the mid-$J_{\rm up}$ (CO(3--2) or CO(4--3)) spectra range from 5 to 20 with a median of 7.
The S/N, FWHMs, line fluxes, and line luminosities of all CO observations we study, including those for CO(1--0), are given in Table~\ref{tab:fluxes}.

\begin{figure}
\centering
	\includegraphics[width=1\columnwidth]{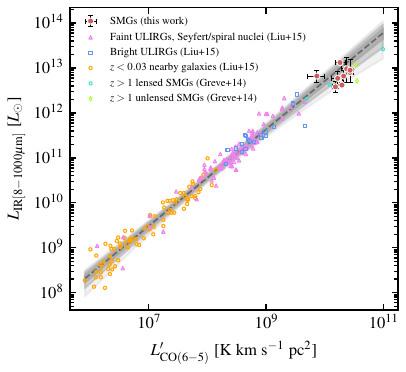}
    \caption{Infrared luminosity ($L_{\rm IR}$) in units of $L_\odot$ versus CO(6--5) line luminosity ($L^{\prime}_{\rm CO,J}$) in units of K km s$^{-1}$ pc$^2$. We compare the relation for our sample against comparison ($L_{\rm FIR}$) samples, for faint ULIRGs, the nuclei of local Seyfert/spiral galaxies, and spatially resolved regions of nearby galaxies from \citet{Liu_2015}, the brightest local ULIRGs from the HerCULES sample \citep{van_der_Werf_2010}, and $z > 1$ SMGs from \citet{Greve_2014}, converting to the same $L_{{\rm IR}[8-1000\upmu {\rm m}]}$ definition as ours by increasing $L_{{\rm FIR}[40-400\upmu {\rm m}]}$ and $L_{{\rm FIR}[50-300\upmu {\rm m}]}$ by factors of 1.05 and 1.12, respectively.
    A linear best-fit line to their (log) measurements is indicated by the dashed line with 1-, 2-, and 3-$\sigma$ confidence levels as increasingly-lighter shading. Our results are consistent with other studies of both local and $z > 1$ systems to within $2\sigma$.}
    \label{fig:lco_lir}
\end{figure}


To investigate the basic high-$J_{\rm up}$ properties of the sample, we calculate line luminosities $L^{\prime}_{\rm CO,J}$ following \citet{Solomon_2005} in units of K km s$^{-1}$ pc$^2$ and investigate the relationship between dust continuum luminosity and CO(6--5) line luminosity -- which acts as a proxy to the Kennicutt-Schmidt (K-S) relation, relating the star-formation rate surface density to gas surface density -- for our sample to provide constraints on the linearity of the relationship with additional high-redshift galaxies, using existing measurements of the infrared luminosities.
We show the relationship in Figure~\ref{fig:lco_lir} for our sample SMGs alongside local sources taken from \citet{Liu_2015}. These include local faint ULIRGs and nuclei of Seyfert and spiral galaxies, spatially resolved regions of nearby galaxies, and the brightest non-extended ULIRGs from the HerCULES sample \citep{van_der_Werf_2010}. We also show measurements of other $z > 1$ SMGs, lensed and unlensed \citep{Greve_2014}.
We calculated the $L_{\rm IR}/L^{\prime}_{\rm CO(6-5)}$ ratio
for each population, finding median and bootstrap errors: for $z < 0.03$ galaxies, $290 \pm 20$;
for ULIRGs, $420 \pm 20$; for $z > 1$ SMGs, $300 \pm 40$; and for our SMGs, $320 \pm 60$.
An orthogonal distance regression (ODR) of the form $\log_{10}(L_{\rm IR}) = N \times \log_{10}(L^{\prime}_{\rm CO(6-5)}) + A$ was applied to the literature values, resulting in best-fit values $N = 1.07 \pm 0.01$ and $A = 1.97 \pm 0.11$. The results of our sample are consistent with the slope and zero-point of local systems to within $2\sigma$ and agree well with other (unlensed) SMGs at $z > 1$.
When we include the results for our SMGs, fixing the slope to the optimum value measured for the literature, we find a zero-point of $A^\prime = 1.77 \pm 0.01$, consistent within $2\sigma$.

\begin{table}
\centering
    \caption{Target ALMA ID, transitions involved in the ratio, and line flux ratio ($R_{i,j}$). Lower limits are due to the $2\sigma$ upper limit CO(1--0) detections by \citet{FriasCastillo_2023} and Jansen et al. (2024, in prep.).}
    \begin{tabular}{@{}ccc@{}}
	\hline\hline
	Target & $^{12}$CO transition ratio & $R_{i,j}$ \\
         & ($i\rightarrow i$--1 $/ j\rightarrow j$--1) & ($I_{\rm CO,i} / I_{\rm CO,j}$) \\
         \hline
        AS2UDS009.0  & $6 \rightarrow 5$ / $1 \rightarrow 0$ & $15.73 \pm 10.58$ \\
                     & $6 \rightarrow 5$ / $3 \rightarrow 2$ & $1.61 \pm 0.36$   \\
        AS2UDS011.0  & $6 \rightarrow 5$ / $1 \rightarrow 0$ & $13.04 \pm 6.13$  \\
                     & $6 \rightarrow 5$ / $4 \rightarrow 3$ & $1.78 \pm 0.37$   \\
        AS2UDS012.0  & $6 \rightarrow 5$ / $1 \rightarrow 0$ & $8.08 \pm 2.74$   \\
                     & $6 \rightarrow 5$ / $3 \rightarrow 2$ & $1.59 \pm 0.35$   \\
        AS2UDS014.0  & $6 \rightarrow 5$ / $1 \rightarrow 0$ & $> 13.25$         \\
                     & $6 \rightarrow 5$ / $4 \rightarrow 3$ & $0.86 \pm 0.17$   \\
        AS2UDS026.0  & $6 \rightarrow 5$ / $1 \rightarrow 0$ & $6.52 \pm 2.76$   \\
                     & $6 \rightarrow 5$ / $4 \rightarrow 3$ & $0.92 \pm 0.24$   \\
        AS2UDS072.0  & $8 \rightarrow 7$ / $1 \rightarrow 0$ & -         \\
                     & $8 \rightarrow 7$ / $4 \rightarrow 3$ & $1.42 \pm 0.31$   \\
        AS2UDS126.0  & $6 \rightarrow 5$ / $1 \rightarrow 0$ & $13.31 \pm 7.65$ \\
                     & $6 \rightarrow 5$ / $3 \rightarrow 2$ & $1.91 \pm 0.43$   \\
        AS2COS0009.1 & $6 \rightarrow 5$ / $1 \rightarrow 0$ & $> 18.38$         \\
                     & $6 \rightarrow 5$ / $3 \rightarrow 2$ & $1.67 \pm 0.33$   \\
        AS2COS0014.1 & $6 \rightarrow 5$ / $1 \rightarrow 0$ & $7.77 \pm 3.08$   \\
                     & $6 \rightarrow 5$ / $3 \rightarrow 2$ & $1.13 \pm 0.18$   \\
        AS2COS0044.1 & $6 \rightarrow 5$ / $1 \rightarrow 0$ & $> 4.83$          \\
                     & $6 \rightarrow 5$ / $3 \rightarrow 2$ & $1.06 \pm 0.36$   \\
        AS2COS0065.1 & $6 \rightarrow 5$ / $1 \rightarrow 0$ & $36.65 \pm 55.01$ \\
                     & $6 \rightarrow 5$ / $3 \rightarrow 2$ & $2.72 \pm 0.22$   \\
        AS2COS0139.1 & $8 \rightarrow 7$ / $1 \rightarrow 0$ & $15.91 \pm 8.27$  \\
                     & $8 \rightarrow 7$ / $4 \rightarrow 3$ & $1.96 \pm 0.31$   \\
	\hline\hline
	\end{tabular}
    \label{tab:line_ratios}
\end{table}


\begin{table}
\centering
    \caption{Summary of measured line ratios for the sample. Columns list the transitions involved in the ratio, the number of sources used to calculate the line ratio, the median spectroscopic redshift of sources used, the median and bootstrap error (including limits) velocity-integrated line flux ratio ($R_{i,j}$), and the median and bootstrap error (including limits) line luminosity ratio ($r_{i,j}$) where the line luminosity was calculated in units of K km s$^{-1}$ pc$^2$.}
    \begin{tabular}{@{}ccccc@{}}
	\hline\hline
	$^{12}$CO transition ratio & N & $z$ & $R_{i,j}$ & $r_{i,j}$ \\
        ($i\rightarrow i$--1 $/ j\rightarrow j$--1) & & & ($I_{\rm CO,i} / I_{\rm CO,j}$) & ($L^{\prime}_{\rm CO,i} / L^{\prime}_{\rm CO,j}$) \\
        \hline
        $3 \rightarrow 2$ / $1 \rightarrow 0$ & 7  & 2.5 & $6.96 \pm 1.83$ & $0.77 \pm 0.21$ \\
        $6 \rightarrow 5$ / $1 \rightarrow 0$ & 10 & 2.8 & $13.14 \pm 2.43$ & $0.37 \pm 0.07$ \\
        $6 \rightarrow 5$ / $3 \rightarrow 2$ & 7  & 2.5 & $1.61 \pm 0.24$  & $0.40 \pm 0.05$ \\
        $6 \rightarrow 5$ / $4 \rightarrow 3$ & 3  & 3.8 & $0.92 \pm 0.39$  & $0.41 \pm 0.17$ \\
	\hline\hline
	\end{tabular}
    \label{tab:median_ratios}
\end{table}

To measure the basic SLED properties, we measure line luminosity ratios ($r_{i,j} = L^{\prime}_{\rm CO(i\rightarrow i-1)}/L^{\prime}_{\rm CO(j\rightarrow j-1)}$) of $r_{6,3} = 0.40 \pm 0.05$ (for 7 sources) and $r_{6,4} = 0.41 \pm 0.17$ (for 3 sources).
The list of all measured line ratios can be found in Table~\ref{tab:line_ratios} and a summary of the median results for the sample can be found in Table~\ref{tab:median_ratios}.
To put these findings into context, the line luminosity ratios of Arp 220, a nearby ultra-luminous infrared galaxy, have been measured as $r_{6,3} = 0.21 \pm 0.06$ and $r_{6,4} = 0.40 \pm 0.03$ \citep{Papadopoulos_2010,Rangwala_2011}.
For the local starburst galaxy M82, $r_{6,3} = 0.50 \pm 0.15$ and $r_{6,4} = 0.67 \pm 0.22$ \citep{Weiss_2005,Ward_2003}.
For SMM J2135-0102 \citep[Cosmic Eyelash,][]{Swinbank_2010}, an archetypal SMG at $z = 2.3$, values have been well constrained as $r_{6,3} = 0.41 \pm 0.02$ and $r_{6,4} = 0.55 \pm 0.04$ \citep{Swinbank_2011,Danielson_2011}.
Measurements by \citet{Weiss_2005} of the gravitationally lensed SMG SMM J16359+6612 at a similar redshift ($z = 2.5$) to our sample find $r_{6,3} = 0.34 \pm 0.05$ and $r_{6,4} = 0.44 \pm 0.11$, consistent with our own.
At higher redshifts, \citet{Spilker_2014} find $r_{6,3} = 0.53 \pm 0.12$ and $r_{6,4} = 0.76 \pm 0.17$, both above our estimates, for their sample of strongly lensed $S_{\rm 1.4mm} \gtrsim$\,20 mJy and $z =$\,2--6 dusty, star-forming galaxies.

Similar multi-$J_{\rm up}$ studies have been carried out using large samples of SMGs, though they typically do not all have multiple lines covered and hence ratios are estimated from the average line luminosities at each transition, statistically corrected for the typical $L_{\rm IR}$ of each sample. For example, \citet{Birkin_2021} estimate $r_{6,3} = 0.43 \pm 0.14$, consistent with our own measurement, and $r_{6,4} = 0.79 \pm 0.23$ which is consistent with our finding. Their analysis was carried out on a sample of fifty SMGs, at the same median redshift of our sample but with a larger range covered ($z =$\,0.7--5), and with $S_{870\upmu{\rm m}} =$\,3--14 mJy.
Likewise, \citet{Bothwell_2013} estimated  $r_{6,3} = 0.40 \pm 0.10$ and $r_{6,4} = 0.51 \pm 0.13$ from their sample of forty luminous ($S_{850\upmu{\rm m}} =$\,4--20 mJy) SMGs at $z =$\,1--4 and \citet{Harrington_2021} estimated $r_{6,3} = 0.36 \pm 0.21$ and $r_{6,4} = 0.48 \pm 0.30$ for a sample of twenty-four strongly lensed DSFGs at $z = $\,1.1--3.5. Both are in reasonable agreement with our measurements.

We find a low-$J_{\rm up}$ line luminosity ratio of $r_{3,1} = 0.77 \pm 0.21$ and high-$J_{\rm up}$ ratio of $r_{6,3} = 0.40 \pm 0.05$ (Table~\ref{tab:median_ratios}; see also \citealt{FriasCastillo_2023}). Although our measurements are higher than those from \citet{Ivison_2011} ($r_{3,1} = 0.55 \pm 0.05$ and $r_{6,3} \sim 0.3$) for their sample of four $z = $\,2.2--2.5 SMGs, they are consistent within the uncertainties.
Our results suggest that our sample of SMGs contains large low excitation gas reservoirs, with slightly lower contributions to the CO(1--0) by their extended cold gas reservoirs than the SMGs studied by \citet{Ivison_2011}.
This is in contrast to sources such as the nuclei of nearby LIRGs which typically exhibit low $r_{3,1}$ but high $r_{6,3}$ because of the potentially larger contribution to the CO(1--0) emission from the extended reservoir \citep{Yao_2003,Leech_2010}.



CO SLEDs can be degenerate in temperature and density at least up to a $J_{\rm up} = 3$ \citep{Bayet_2006,Papadopoulos_2008,Dannerbauer_2009,Leech_2010,Papadopoulos_2012}.
Without constraints on the average line optical depths, which would be potentially obtainable using $^{13}$CO observations, temperature and density degeneracies remain at $J_{\rm up} \geq 3$.
For these reasons, estimates of the molecular gas mass have typically been found from measurements of the CO emission of transitions up to $J_{\rm up} = 3$, assuming the transitions below and including $J_{\rm up} = 3$ are close to thermalised.
However, as reported by \citet{FriasCastillo_2023}, $J_{\rm up} = 3$ line intensities for the sample normalised by CO(1--0) demonstrate significant scatter, suggesting that the assumption of a constant low-$J_{\rm up}$ ratio is not entirely valid in SMGs of $z = $\,2--4.






\subsubsection{Dust emissivity}

Besides CO SLEDs, another useful indicator of the physical and chemical properties of the ISM is the dust emissivity spectral index, $\beta$, describing the frequency dependence of the dust emissivity per unit mass. 
The value of $\beta$ for DSFGs has typically been assumed to be in the range $\beta = $\,1--2, while recent studies suggest that it may be even higher \citep[e.g.,][]{Casey_2021,daCunha_2021,Cooper_2022}. The evolution of the dust emissivity index with redshift is also a subject of some studies \citep[e.g.,][]{Ismail_2023,Witstok_2023}, with \citet{Ward_2024} reporting no evidence for such a trend and attributing variations of $\beta$ to intrinsic variations in the properties of the dust in DSFGs.

We measured $\beta$ for the SMGs in our sample, using the continuum observations from our Band 5/6 spectra combined with the published $870\upmu$m flux derivatives from \citet{Simpson_2020} and \citet{Stach_2019}, and found a median and bootstrap error of $\beta = 1.4 \pm 0.5$ (individual measurements can be found in Table~\ref{tab:line_ratios}). Our measurement is consistent with the emissivity index of dust in the Milky Way and other local and high-redshift galaxies ($\beta \sim $\,1--2), and also consistent with \citet{daCunha_2021} from 2mm observations of 27 SMGs who measured $\beta = 1.9 \pm 0.4$.

\subsection{CO SLEDs}
\label{sec:sleds}

\begin{figure*}
\centering
	\includegraphics[width=1\linewidth]{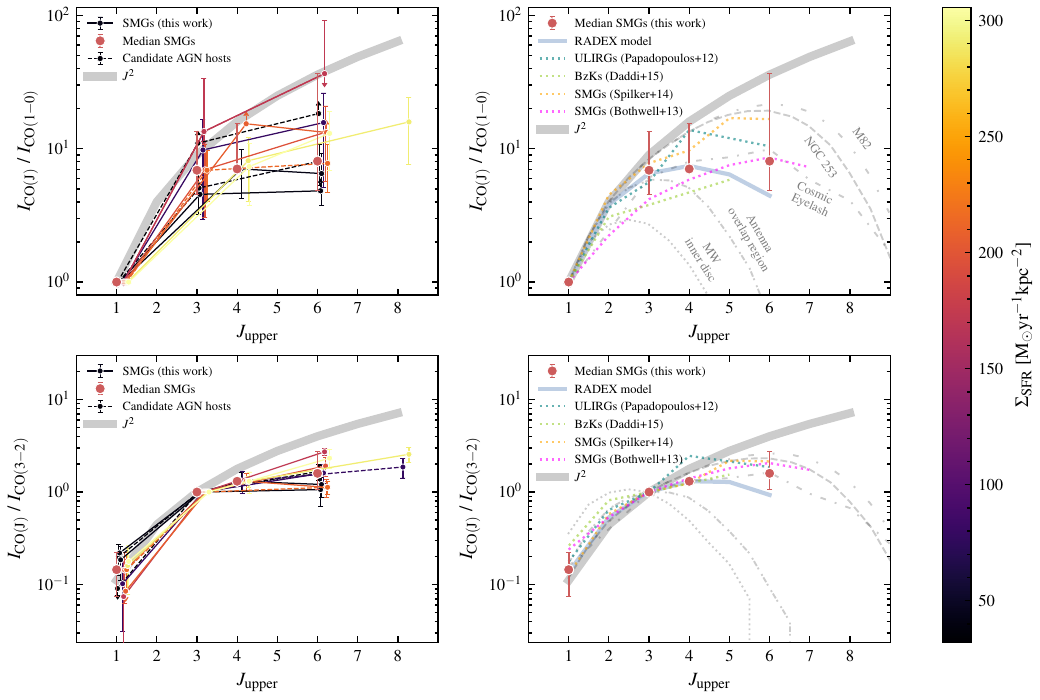}
    \caption{CO SLEDs for our sample SMGs. Individual fluxes measured from fits to the spectra are shown as circles with error-bars indicating their propagated uncertainties determined via a Monte Carlo method.
    {\it Top left:} Fluxes are normalised to the flux (or upper limits) of the CO(1--0) line and are coloured by their respective star-formation rate surface densities denoted by the colour bar (applicable to both top- and bottom-left panels). AS2COS0065.1 is shown with a limit for the lower error since it causes the error-bar to decrease below a line ratio of 1. We also show the combined composite SLED of the sample (circles), normalised by the star-formation rate surface density ($\Sigma_{\rm SFR}$) and 68\% confidence level as error-bars.
    {\it Top right:} The combined composite SLED of the sample (circles) with errors corresponding to the scatter in the sample, similar to systems such as SMM J2135-0102 (Cosmic Eyelash), NGC 253, and local ULIRGs. $J_{\rm up} = 8$ measurements are omitted for the median, since there is only one available.
    Shown alongside is the best-fit \texttt{RADEX} model to the composite CO SLED, computed from a grid of temperatures and densities, which suggests average ISM conditions of $T_{\rm kin} \sim $\,30--50~K, $\log_{10}(\langle n_{\rm H_2}\rangle [{\rm cm}^{-3}]) \sim $\,3.2--3.5, and $\langle dv/dr \rangle = $\,0.5--5\,kms$^{-1}$pc$^{-1}$, though there is large covariance.
    {\it Bottom left:} Similar to the top-left panel, but fluxes are normalised to the flux of the CO(3--2) where they are available, or are otherwise normalised by the predicted flux of the CO(3--2) from the CO(4--3) using the line/brightness temperature ratio measured from \citet{Danielson_2011}.
    {\it Bottom right:} Similar to the top right panel, but with fluxes normalised to the CO(3--2).
    We add a small offset to the $J_{\rm up}$ of each measurement for each source to separate their error-bars which would otherwise overlap.
    We also show $J_{\rm up}^2$ scaling which is the expected scaling of intensities for levels in LTE and in the Rayleigh-Jeans limit.
    We compare to the CO SLEDs of other systems such as the Milky Way inner disc \citep{Fixsen_1999}, the overlap region of the antennae galaxies NGC 4038 and NGC 4039 \citep[Antenna overlap region;][]{Mashian_2015}, SMM J2135-0102 \citep[Cosmic Eyelash;][]{Danielson_2011}, M82 \citep{Weiss_2005}, NGC 253 \citep{Israel_1995,Harrison_1999,Israel_2002,Bradford_2003}, ULIRGs \citep{Papadopoulos_2012}, BzKs \citep{Daddi_2015}, and SMGs from \citet{Spilker_2014} as well as \citet{Bothwell_2013}.
    }
    \label{fig:sled}
\end{figure*}

In Figure~\ref{fig:sled}, we plot the CO SLEDs of our sample of SMGs.
We show the SLEDs of all the sources, normalised to the CO(1--0). These rise towards $J_{\rm up} = 6$ indicating intense star-forming conditions of their ISMs, although, because we do not have measurements of the $J_{\rm up} = 5$ or $J_{\rm up} \geq 7$, we cannot conclude that the SLEDs peak at $J_{\rm up} = 6$.
There is a large diversity in the excitations of the SLEDs even at $J_{\rm up} = 3$ with respect to the CO(1--0), by a factor of $\approx 2.5$ (given the errors dominated by the low S/N CO(1--0) measurements). At $J_{\rm up} = 4$, we measure a scatter of a factor of $\approx 1.5$ and at $J_{\rm up} = 6$ a scatter of $\approx 3$.
We also show the line intensities normalised by the CO(3--2) which provides us with better constraints on the diversity of excitation of the star-forming-only gas, since the CO(3--2) was in general detected to much higher S/N than the CO(1--0). For sources that do not have available CO(3--2) measurements, we predict them from CO(4--3) using the line/brightness temperature ratio measured for the SMG SMM J2135-0102 \citep{Danielson_2011}.
Here, it is clear by the scatter of line ratios between CO(3--2) and CO(6--5) of a factor of $\approx 2$ that the diversity of excitation found at high-$J_{\rm up}$ is significant.
Where the fluxes are normalised to the CO(1--0) luminosity, there is a qualitative trend between increased excitation of the CO SLED and $\Sigma_{\rm SFR}$ indicated by the colouring of the SLEDs, which is well-demonstrated at $J_{\rm up} = 6$ by the scatter in excitation of a factor of $\approx 3$.
Individual SLEDs normalised by the CO(4--3) can also be found in Figure~\ref{fig:extra_sled}.


In addition, we show the $J_{\rm up}^2$ scaling which is the expected scaling of intensities for energy levels of optically thick and thermalised gas in local thermodynamic equilibrium (LTE) and in the Rayleigh-Jeans limit.
Under these conditions, the excitation temperature equals the kinetic temperature of the gas, and the intensities of certain emission lines become predominantly determined by the gas temperature, via the Boltzmann distribution, rather than density.
Sources with global velocity-integrated flux density ratios exceeding the $J^2$ scaling would have corresponding brightness temperature ratios $> 1$. Such ratios, however, caused by LTE-excited and optically thin CO line emission dominating the global H$_2$ mass reservoirs, are very rarely observed for galaxy-sized gas reservoirs (e.g., NGC 3310, \citealt{Zhu_2009}).
None of our sources have significant excitation above $J_{\rm up}^2$ scaling.

The sources we consider as potential AGN hosts are indicated in Figure~\ref{fig:sled} with dashed lines between their line ratios. There is clearly a large diversity in the shapes of their SLEDs at $J_{\rm up} \geq 3$.
Although it is low S/N, it is possible that the $J_{\rm up} = $\,3 CO(1--0)-normalised $2\sigma$ lower limit line ratio of AS2COS0009.1 that exceeds $J^2$ could be due to an AGN that may be heating sufficiently large molecular mass fractions to high kinetic temperatures and that the gas is kinematically violent enough to cause optically thin CO line emission.
In Section~\ref{sec:drivers}, we test the relationship between the CO SLED shape and the potential physical drivers using the excitation between CO(3--2) and CO(6--5), which can be understood to be independent of any effects that hosting an AGN may have on the ISM of an SMG. Therefore, we can be confident that our observations provide a reliable test against the predictions from the \citetalias{Narayanan_2014} model.

We determined the composite SLED of the sample by computing the median of line ratios normalised by the star-formation rate surface density of each source, under the assumption that it correlates strongly with CO excitation. $J_{\rm up} = 8$ measurements are omitted, however, since there is only a maximum of two available sources.
We also test how the combined SLED of the sample changes if we normalise by infrared luminosity and we recover the same result albeit with larger scatter at $J_{\rm up} = 3$.
We show the composite SLED in Figure~\ref{fig:sled}, with the SLEDs of other systems for comparison.
When normalised by CO(1--0), the median SLED of our sample clearly increases above and beyond that of the Milky Way inner disc (a factor $\approx 3$ greater at $J_{\rm up} = 3$) which peaks at around $J_{\rm up} = 2$, and follows excitations similar to NGC 253 and local ULIRGs \citep{Bradford_2003,Papadopoulos_2012}. Unfortunately, the lack of $J_{\rm up} = 5$ and $J_{\rm up} \geq 7$ lines means that we are unable to unambiguously identify the peak of the SLED. For this reason, we cannot distinguish which specific class of ULIRGs characterised by \citet{Rosenberg_2015} the SLED is most similar to.
Like local ULIRGs, observations suggest that SMGs may be triggered by mergers or interactions \citep{Ivison_2007,Tacconi_2008,Engel_2010,Alaghband_2012,Chen_2015}, though simulations suggest that most may be driven by the hosting of an undermassive black hole that maintains an adequate gas reservoir \citep{McAlpine_2019}. SMGs generally have more extended dust continuum sizes \citep[$\gtrsim 2$\,kpc; ][]{Swinbank_2012,Simpson_2019,Gullberg_2019} and higher dust continuum luminosities which appear to be powered by star formation \citep{Pope_2006,Pope_2008}, compared to more compact ($\approx 1$\,kpc), lower luminosities in local ULIRGs powered primarily by starbursts and in some cases a central AGN \citep[e.g.,][]{Genzel_1998}.
A likeness to the CO SLEDs of local ULIRGs has been hypothesised as the result of a decline in metallicity toward high redshift, which may cause the intensity of the radiation field $\langle U \rangle$ in SMGs to become similar to local ULIRGs \citep{Daddi_2015}.
Although the $J_{\rm up} = $\,3~and~4 excitations of the median SLED are more highly excited than other SMGs selected with $S_{870} = $\,4--20\,mJy at $z \approx $\,1--4 \citep{Bothwell_2013}, it is consistent with other bright SMGs ($S_{870} \gtrsim $\,20~mJy) at $z \approx $\,2--6 \citep{Spilker_2014}.

We also show the best-fit temperature-density \texttt{RADEX}
\citep{vanderTak_2007} model to the composite SLED, computed from the
interpolated temperature-density grid taken from \citet{Liu_2021b} for
local and high-redshift ($z \sim $\,0--6) galaxies based on $J_{\rm
up} = 2$ and $J_{\rm up} = 5$ CO line measurements.
Since \texttt{RADEX} does not consider the velocity gradients
($dv/dr$) likely present in macroturbulent GMCs that are likely within
the SMGs, we use the model to determine the column density, and
consequently $\langle dv/dr \rangle$, which could describe the SLED,
using the optimal grid $T_{\rm kin}$ and $n_{\rm H_2}$ values as
inputs, and CO abundance $\bigl[{\rm CO}/{\rm H}_2\bigr] = 5 \times
10^{-5}$ consistent with \citet{Liu_2021b}.
There is strong covariance between temperature and density, meaning that on average the SMGs can be described simultaneously with high density and low temperature or low density and high temperature, with solutions that include velocity gradients as low as $dv/dr \sim 0.1$\,kms$^{-1}$pc$^{-1}$.
However, on average, the clouds are likely to be virial (or
super-virial).  We therefore use the approximation by
\citet{Davies_2012} to provide an estimate of the likely velocity
gradient for the virialised case (which may be considered a reasonable
physical lower limit) and obtain $dv/dr \sim 0.5$\,kms$^{-1}$pc$^{-1}$.  For
the family of best-fit models to the $\Sigma_{\rm SFR}$-normalised
composite SLED which are higher than $dv/dr \sim
0.5$\,kms$^{-1}$pc$^{-1}$, the typical ISM can be described with
conditions of $T_{\rm kin} \sim $\,30--50~K and $\log_{10}(\langle
n_{\rm H_2}\rangle [{\rm cm}^{-3}]) \sim $\,3.2--3.5 (with a maximum
velocity gradient up to $dv/dr \sim 5$\,kms$^{-1}$pc$^{-1}$, but with
degeneracies between temperature and density).  Nevertheless, these ranges are similar to the physical conditions of the gas in the cores of local
starburst galaxies such as M82 and Arp 220 \citep{Weiss_2005,Rangwala_2011}.

\subsection{Drivers of CO excitation}
\label{sec:drivers}

Constraints on the CO spectral line energy distributions (SLEDs) of galaxies reflect the physical conditions of the star-forming regions that control their star formation. Determining the optimal parameterisation for their shapes would enable the prediction of lower transition fluxes and subsequent molecular gas mass measurements to better accuracies and precisions, since they can be converted using suitable CO-to-H$_2$ conversion factors.

The lack of a unique template CO SLED describing any particular galaxy population means, in the absence of direct measurements of low-$J_{\rm up}$ ($J_{\rm up} \leq 3$) emission which are observationally expensive for large samples, we need a robust method to determine the appropriate SLED to infer low-$J_{\rm up}$ luminosities and hence the total molecular gas masses of individual galaxies.
To address this problem, \citetalias{Narayanan_2014} proposed a simple parameterised model for the CO SLED based on the expectation that the CO excitation is primarily dependent on the gas temperatures, densities, and optical depths in the molecular ISM. They used {\sc gadget}-3 to simulate a range of galaxy structures, including regular discs and a variety of mergers covering a range of masses, merger orbits, and halo virial properties, and then used the properties of the SPH gas particles from these, along with a physical model for the thermal and physical structure of the star-forming ISM from \citet{Krumholz_2014}, to provide a parameterisation of CO SLEDs in terms of the star-formation rate surface density ($\Sigma_{\rm SFR}$).
In their model, the shape of CO SLEDs between $J_{\rm up} = 3$\,and~6 correlates well with the resolved and galaxy-averaged $\Sigma_{\rm SFR}$, and it was able to generally reproduce the SLEDs of fifteen star-forming galaxies from $z > 6$ to the present epoch.
\citet{Daddi_2015} found their sample of four $z = 1.5$ star-forming disc galaxies to have an average CO SLED that disagrees with the shape predicted by the \citetalias{Narayanan_2014} model, with a smaller flux ratio $R_{3,1}$ than expected, and concluded that the primary driver of the SLED shape is the strength of the interstellar radiation field, $\langle U \rangle$, reflected in dust temperature \citep{Magdis_2012}, and which is directly
related to the $\Sigma_{\rm SFR}$.
\citet{Wu_2015}, found similar results while studying resolved observations of the star-forming regions in M83, attributing excitation to mechanical heating induced by recent nuclear starbursts.
In fact, the results by \citet{Daddi_2015} agree better with the model proposed by \citet{Papadopoulos_2012} -- which predicts that the SLED should continue to rise up to $J_{\rm up} = 5$ and
that the gas responsible for the high-$J_{\rm up}$ CO line emission is heated by turbulence and/or cosmic rays
-- since there is no evidence that the SLED has reached its peak even at $J_{\rm up} = 5$.
Similarly, \citet{Valentino_2020} found that while the CO excitation in $z = $\,1--1.7 main-sequence and starburst galaxies generally correlates with $\langle U \rangle$, it more closely correlates with $\Sigma_{\rm SFR}$ and, most significantly, with SFE, though they conclude that $\Sigma_{\rm SFR}$ better captures the gas temperatures and densities.

In this section, we test the various model predictions which attempt to describe the shape of the SLED using our sample SMGs, focusing on the model proposed by \citetalias{Narayanan_2014} which considers the main driver of excitation to be star-formation rate surface density ($\Sigma_{\rm SFR}$).
To investigate the correlation between CO excitation and physical drivers, first, we use the $I_{\rm CO(6-5)}/I_{\rm CO(3-2)}$ ($R_{6,3}$) line ratio since we have measurements of the CO(3-2) line flux for the majority of (7 out of 12) sources in the sample. For four sources, the mid-$J_{\rm up}$ transition is from $J = $\,4--3 and, for consistency in the following analysis, we again use the CO(3--2) values predicted from the adoption of the line ratio for SMM J2135-0102 \citep[Cosmic Eyelash; ][]{Danielson_2011}, in the same way as in the bottom-left panel of Figure~\ref{fig:sled}.

\begin{figure*}
\centering
	\includegraphics[width=0.85\textwidth]{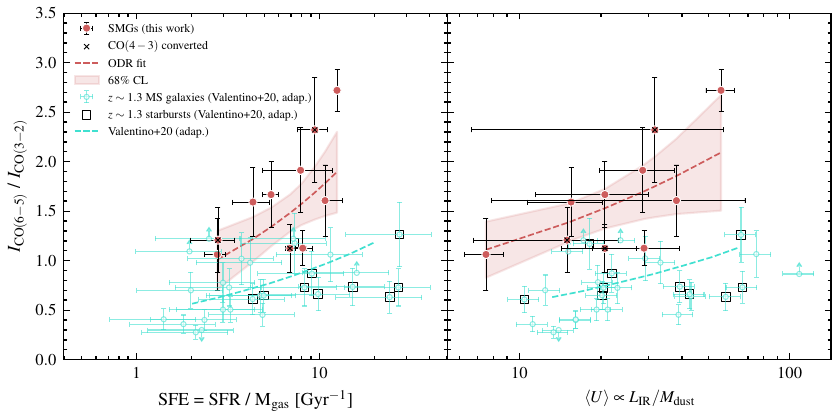}
    \caption{$^{12}$CO SLED shapes of ten sources from our sample of SMGs with CO(6--5) observations, probed by measurements of their $J_{\rm up} = 3$ and $J_{\rm up} = 6$ CO line fluxes, versus proposed drivers of excitation, SFE ({\it left}) and $\langle U \rangle$ ({\it right}). For sources without a CO(3--2) measurement, we derive the predicted flux from the CO(4--3) measurement using the line/brightness temperature ratio measured for the archetypal SMG SMM J2135-0102 \citep{Danielson_2011}. We exclude sources AS2UDS072.0 and AS2COS0139.1, however, since the available high-$J_{\rm up}$ line was the CO(8--7) and the same excitation processes may not be present in both $J_{\rm up} = 6$ and $J_{\rm up} = 8$ regions of emission.
    We show orthogonal distance regression (ODR) fits to our observations with the 68\% confidence levels estimated by bootstrapping.
    We also compare our results to those from \citet{Valentino_2020} for $z \sim $\,1.3 main-sequence galaxies and starbursts, where we converted their $R_{5,2}$ line ratios to $R_{6,3}$ using line flux ratios taken from \citet{Weiss_2005} for local starburst M82, with typical error-bars associated with the data.
    Correlation coefficients and probabilities for the relationships can be found in Table~\ref{tab:MIC}.
    Offsets between the line ratios of SMGs and low-redshift galaxies at fixed SFE or $\langle U\rangle$ indicate that these drivers may not be optimal to describe CO excitation, whereas models improve by considering galaxy sizes (Figure~\ref{fig:model}).
    }
    \label{fig:sfe_isrf_models}
\end{figure*}


\subsubsection{The star-formation efficiency and the average intensity of the interstellar radiation field}


An increase in the efficiency of star formation, caused by an enhancement of the fragmentation in gas-rich, turbulent, and gravitationally-unstable high-redshift discs \citep{Bournaud_2007,Dekel_2009,Ceverino_2010,Dekel_2014}, reflected in their clumpy morphologies \citep{Elmegreen_2007,Forster_2011,Genzel_2011,Guo_2012,Zanella_2019}, would correspond to higher star-formation rates on the assumption that the gas mass stays constant.
However, recent works suggest that increased star-formation rates are accompanied by increased gas masses \citep{Scoville_2016,Elbaz_2018}. Mergers could explain the accumulation of new gas, and are likely the case in the local Universe (see \citealt{Lonsdale_2006} and references therein), however some SMGs at high redshifts have also been shown to host orderly rotating discs \citep{Hodge_2016,Hodge_2019,Drew_2020,Birkin_2024} countering this idea.
The results from \citet{Valentino_2020} suggested that the strongest correlator against CO excitation in low-redshift star-forming galaxies, at least when using the $R_{5,2}$ line ratio as a proxy, is star-formation efficiency. For this reason, we investigated the relationship between CO excitation and star-formation efficiency in our sample SMGs.

To derive star-formation efficiency, we convert the CO(1--0) line luminosities of the SMGs measured by \citet{FriasCastillo_2023} to molecular gas masses assuming a CO-to-H$_2$ conversion factor $\alpha_{\rm CO} = 1$, common for ULIRGs\footnote{ULIRGs have also been found with as high as $\alpha_{\rm CO} = 5$ \citep{Dunne_2022} but are not consistent with resolved dynamics of SMGs \citep{Rivera_2018,Amvrosiadis_2023}.} and starburst galaxies, with a 1.36 factor to account for Helium abundance \citep{Bolatto_2013}. It should be noted that our definition of star-formation efficiency differs from that used in \citet{Valentino_2020} which converted dust masses to gas masses using a metallicity dependent gas-to-dust ratio \citep{Magdis_2012}. The SMGs have a range of ${\rm SFE} = $\,2.6--12.5\,Gyr$^{-1}$ with a median ${\rm SFE} = $\,7.4\,Gyr$^{-1}$.

The average intensity of the radiation field $\langle U\rangle$ can be expected to relate to star-formation efficiency through an assumption of the metallicity \citep[$\langle U\rangle \propto {\rm SFE}/Z$;][]{Magdis_2012}. 
It could therefore be a useful component to distinguish galaxy populations in a way that reflects differences in the excitation of the ISM.
Moreover, we expect harder radiation fields in dense regions of the ISM which, in combination with increased cosmic ray rates, heats the dust. Inherently, this suggests that $\langle U\rangle$ may not necessarily be a {\it driver}, but a symptom, of CO excitation.
\citet{Daddi_2015} suggested that the main driver of $R_{5,2}$ CO excitation in low-redshift star-forming galaxies is $\langle U\rangle$, and that it is not closely correlated with star-formation efficiency. We therefore also investigate the relationship between CO excitation and $\langle U\rangle$ in our sample SMGs.

We adopt the same definition for the average intensity of the radiation field $\langle U\rangle = 1/P_0 \times L_{{\rm IR}[8-1000\upmu {\rm m}]} / M_{\rm dust}$ as \citet{Valentino_2020}, where $P_0 = 125$\,$L_\odot/M_\odot$ so that $\langle U\rangle$ is dimensionless \citep{Draine_2007}, a quantity related to the power absorbed per unit dust mass in a radiation field. The sample galaxies have a range of $\langle U\rangle = $\,7.5--55.7 with a median $\langle U\rangle = $\,28.7.

We show how the CO excitation changes as a function of star-formation efficiency and $\langle U\rangle$ for our sample SMGs in Figure~\ref{fig:sfe_isrf_models}. We also compare our results to those found by \citet{Valentino_2020}, for the $J_{\rm up} = $\,2~and~5 CO line measurements from $z = $\,1.0--1.7 main sequence and starburst galaxies. The parameter space covered by their sample consists of $\langle U\rangle = $\,10--100 and ${\rm SFE} = $\,1--30~Gyr$^{-1}$, similar to our sample.
We converted their $r_{5,2}$ line luminosity ratios to equivalent $R_{6,3}$ line fluxes using $R_{3,2}$ and $R_{6,5}$ line fluxes taken from \citet{Weiss_2005} for the local starburst galaxy M82. We may expect some variations due to uncertain conversion factors, though using line fluxes from \citet{Danielson_2011} for SMM J2135-0102 only decreases the line ratios by 3\%, so these values are not expected to change by significant amounts depending on the system of choice used to convert them.
It is clear that SMGs tend to display larger CO line ratios than main-sequence and starburst galaxies.
At fixed SFE or $\langle U\rangle$, $R_{6,3}$ line ratios of SMGs are as much as 3 or 4 times larger than those of low-redshift galaxies, respectively.
Such large offsets indicate that CO excitation cannot be accurately described by these drivers simultaneously in the ISMs of low-redshift, less active systems and high-redshift, highly star-forming systems. The size encapsulated in the star-formation rate surface density, suggested by the \citetalias{Narayanan_2014} model, is instead expected to provide an optimal parameterisation of the excitation \citep{Puglisi_2021}.

\begin{figure*}
\centering
	\includegraphics[width=0.65\textwidth]{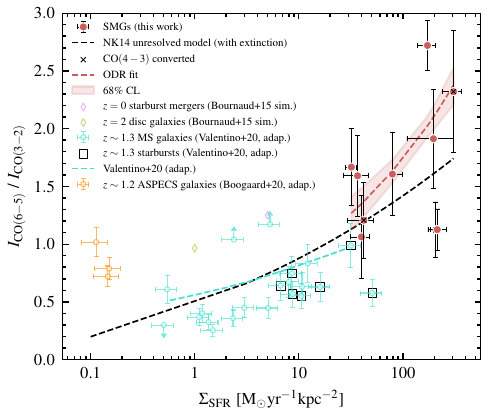}
    \caption{$^{12}$CO SLED shapes of ten sources from our sample of SMGs with CO(6--5) observations, probed by measurements of their $J_{\rm up} = 3$ and $J_{\rm up} = 6$ CO line fluxes, versus star-formation rate surface density ($\Sigma_{\rm SFR}$). For sources without a CO(3--2) measurement, we find the predicted flux from the CO(4--3) measurement using the line/brightness temperature ratio measured for the archetypal SMG SMM J2135-0102 \citep{Danielson_2011}. We exclude sources AS2UDS072.0 and AS2COS0139.1, however, since the available high-$J_{\rm up}$ line was the CO(8--7) and the same excitation processes may not be present in both $J_{\rm up} = 6$ and $J_{\rm up} = 8$ regions of emission.
    We show an orthogonal distance regression (ODR) fit to our observations with the 68\% confidence level estimated by bootstrapping.
    We compare to the unresolved model from \citetalias{Narayanan_2014} with added extinction, which predicts that the increasing SLED shape correlates with sources with higher $\Sigma_{\rm SFR}$, suggesting more intense ISM properties.
    We also compare our results to those from \citet{Bournaud_2015} for simulated $z = 0$ starburst mergers and $z = 2$ discs, as well as \citet{Valentino_2020} for $z \sim$\,1.3 main-sequence galaxies and starbursts and \citet{Boogaard_2020} for $z \sim $\,1.2 ASPECS galaxies, where we converted their $R_{5,2}$ line ratios to $R_{6,3}$ using line flux ratios taken from \citet{Weiss_2005} for local starburst M82, with typical error-bars associated with the data. The fit to the low-redshift measurements was computed on the sources shown with the addition of others that had upper limits on $\Sigma_{\rm SFR}$.
    The correlation coefficient and probability for the relationship can be found in Table~\ref{tab:MIC}.
    The observations of our SMGs are consistent with the \citetalias{Narayanan_2014} model to within $3\sigma$. However, we find significant scatter, indicative of intrinsic variations in the ISM for a given $\Sigma_{\rm SFR}$, which could be described by other physical properties.
    In combination with the observations from \citet{Valentino_2020}, we find that the strongest driver of CO excitation is $\Sigma_{\rm SFR}$.}
    \label{fig:model}
\end{figure*}

\subsubsection{Star-formation rate surface density}
\label{sec:model}

In Figure~\ref{fig:model}, we show the relationship between CO excitation and star-formation rate surface density for our SMGs alongside low-redshift galaxies from \citet{Valentino_2020}. Their sample provides us with an opportunity to test the relationship across nearly three orders of magnitude, since it covers the range $\Sigma_{\rm SFR} = $\,0.5--50\,M$_\odot$yr$^{-1}$pc$^{-2}$.
The average $R_{6,3}$ line ratio for our SMGs is a factor of $\approx 2.5$ times greater than low-redshift galaxies, suggesting that the SLEDs of SMGs may peak at higher $J_{\rm up}$. Our SMGs also appear to be more excited than the extrapolation of the fit to the low-redshift galaxies, indicating that the relationship between excitation and star-formation rate surface density may be different in the two populations.
We also compare our observations to $z \sim $\,1.2 ASPECS galaxies from \citet{Boogaard_2020}, similarly converting their $R_{5,2}$ line ratios to $R_{6,3}$ using line ratios measured for M82 \citep{Danielson_2011}. The three ASPECS galaxies show signs of a negative trend between CO excitation and $\Sigma_{\rm SFR}$, contrary to the predictions from \citetalias{Narayanan_2014}, though the range covered for the physical driver is too small to assign it any significance.

We computed the maximal information coefficient (MIC) using {\sc minepy} \citep{Albanese_2012} to carry out a maximal information-based nonparametric exploration (MINE) of the correlations between CO SLED shape and the possible physical drivers, star-formation efficiency, $\langle U\rangle$, and star-formation rate surface density. Since the parameter space covered by the SMGs in our sample is limited, we combine our measurements with those of low-redshift galaxies from \citet{Valentino_2020}. However, we highlight that the number of sources supplementing our data from their work differs for each driver of CO excitation and we only use detections, discarding lower and upper limits on $R_{6,3}$. In addition, we do not include the ASPECS galaxies from \citet{Boogaard_2020} since they only apply to star-formation rate surface density. To confirm the significance of our results, we also computed the probability that the measured strength of the correlation could be found in the available data, via Jackknife resampling.
The observable which we find to be the strongest driver of CO excitation in star-forming galaxies is $\Sigma_{\rm SFR}$, with an ${\rm MIC} = 0.81$ and $p = 0.96$, significant to $2\sigma$. The results of this test can be found in Table~\ref{tab:MIC}.

\begin{table}
\centering
    \caption{Maximal information coefficients (MIC) and probabilities ($p$), determined from Jackknife resampling, between CO excitation and the possible physical drivers, SFE, $\langle U\rangle$, and $\Sigma_{\rm SFR}$, using the $R_{6,3}$ line ratios of our sample SMGs combined with those of $z = $\,1.0--1.7 main-sequence and starburst galaxies from \citet{Valentino_2020}.
    We find that the strongest driver of CO excitation in star-forming galaxies is $\Sigma_{\rm SFR}$.}
    \begin{tabular}{|c|c|c|}
    \hline\hline 
    Driver of CO excitation & MIC & $p$ \\
    \hline
    SFE & 0.64 & 0.96 \\
    $\langle U\rangle$ & 0.34 & 0.79 \\
    $\Sigma_{\rm SFR}$ & 0.81 & 0.96 \\
    \hline\hline
    \end{tabular}
    \label{tab:MIC}
\end{table}





The model proposed by \citetalias{Narayanan_2014} predicts a relationship between CO excitation and $\Sigma_{\rm SFR}$ that follows a power law.
Though a power-law fit may describe the behaviour expected for ground-state normalised line intensity ratios as a function of transition number in low-$\Sigma_{\rm SFR}$ galaxies, it may not necessarily be an accurate description of the relationship at high-$\Sigma_{\rm SFR}$.

Large dust columns produce high optical depths and so reduce the intensities measured for high-$J_{\rm up}$ lines \citep{Papadopoulos_2010,Rangwala_2011}. The corrections required use a dust extinction model which assumes that the gas and dust are well mixed in the ISM.
The line intensities predicted by the \citetalias{Narayanan_2014} model are the intrinsic luminosities. To make comparisons between the simulation results and our observational measurements, we apply a correction to the simulation.
We add extinction to their model of intrinsic $R_{6,3}$ flux ratios such that 
\begin{equation}
R_{6,3} = R_{6,3}^{\rm int} \times \exp \biggr[ -\tau_d(\nu_{32}) \biggl( \Bigl( \frac{\nu_{65}}{\nu_{32}} \Bigr) ^\beta -1 \biggr) \biggr]
\end{equation}
\noindent where $\tau_{d} = (\nu/\nu_{0})^{\beta}$ and a spectral index of $\beta = 2$ \citep{Tacconi_2018,daCunha_2021,Liao_2024} at a reference frequency $\nu_0 = 353$\,GHz (i.e., $\uplambda_0 = 850$\,$\upmu$m).
This results in a decrease of the $R_{6,3}$ line ratios predicted by the \citetalias{Narayanan_2014} model by an average of 1.7.

We now quantify the $R_{6,3}$--$\Sigma_{\rm SFR}$ trend, first using a linear relation and then using a power law to test the model proposed by \citetalias{Narayanan_2014}.
We applied a linear orthogonal distance regression to the data to account for uncertainties in both $R_{6,3}$ and $\Sigma_{\rm SFR}$, of the form $\log_{10}(R_{6,3}) = m \times \log_{10}(\Sigma_{\rm SFR}) + c$, finding best-fit values and bootstrap errors for the slope $m = 0.28 \pm 0.07$ and for the zero point $c = -0.3 \pm 0.2$, and determined the 68\% confidence level via bootstrapping.
The coefficients of the \citetalias{Narayanan_2014} model are empirical,
but are a consequence of changes to the radiation field, $\langle U \rangle$, and star-formation efficiency in the model, which in turn cause differences in the gas excitation and star-formation rate surface density. To test the parameterisation of the model proposed by \citetalias{Narayanan_2014}, we applied a power-law fit to the data following:
\begin{equation}
    \log_{10}(R_{6,3}) = A \times \left[\log_{10}\left(\Sigma_{\rm SFR}\right)-\chi\right]^B + C
\end{equation}

\noindent where $\chi = -1.85$, used by \citetalias{Narayanan_2014} to ensure the prediction of real values, and found best-fit parameters and bootstrap errors of $A = 0.0 \pm 0.3$, $B = 3.5 \pm 0.9$, and $C = -0.1 \pm 0.4$.

We tested the form of the relationship by applying the Bayesian information criterion (BIC) to the linear and power-law fits, which introduces a penalty term in the goodness of fit test for functions with every additional parameter used. The linear fit provides a better parameterisation of the relationship between excitation and star-formation rate surface density, at least across the range $\Sigma_{\rm SFR} \approx $\,30--300~M$_\odot$~yr$^{-1}$~kpc$^{-2}$.

We show the linear fit to our sample SMGs in Figure~\ref{fig:model}, alongside the \citetalias{Narayanan_2014} unresolved empirical model with added extinction for comparison.
The correlation between the galaxy-integrated CO SLED shape and star-formation rate surface density in the model proposed by \citetalias{Narayanan_2014} provides a reasonable description of the data (when using CO(3--2) and CO(6--5) line fluxes as a proxy for the excitation of the gas), which is within $\sim 3\sigma$.
There is, however, significant scatter of a factor of $\approx 3$ in the line flux ratios that we measure which reflects intrinsic variations in the ISM properties rather than measurement uncertainties.

We note that the line intensity ratios describing the molecular gas excitation may correlate more specifically with the star-formation rate {\it volume} density, rather than the surface density. The model proposed by \citetalias{Narayanan_2014} predicts a relationship between CO excitation and star-formation rate density, which is represented as a surface density for ease of comparison with observations. Thus, the treatment of an observed star-formation rate surface density as a proxy for the volume density should not be made without considering how it is affected by the scale height of the disc. To adopt different scale heights for different sources -- since low-$\Sigma_{\rm SFR}$ SMGs might be more disc-like and high-$\Sigma_{\rm SFR}$ SMGs might have higher velocity dispersions due to mergers seen at random angles -- we would need to assume some distribution of their values. Instead, we consider how the extreme cases change the resulting observed relationship between CO excitation and star-formation rate density. For the smallest scale height expected of high-redshift discs, we assume for simplicity the minimum scale height measured for the thick disc of the Milky Way\footnote{Though this may not be applicable to strong mergers such as Mrk 231 and Arp 220 where star-formation rate densities are huge and within very small volumes ($r \leq 200$\,pc).}, $h_{\rm min} \approx 0.6$\,kpc \citep{Hawthorn_2016}, while the largest scale height would be equal to the dust continuum radius under the assumption of a spherical geometry, $h_{\rm max} \approx 1.3$\,kpc. Adopting the minimum scale height results in an increase in $\Sigma_{\rm SFR}$ of a factor of 1.7, causing the observations to deviate from the \citetalias{Narayanan_2014} model by $\approx 2.5\sigma$, and reduces any systematic offset between them.
Adopting the maximum scale height leads to a decrease in $\Sigma_{\rm SFR}$ of a factor of 1.6 which causes the observations and model to differ by $\approx 3\sigma$. Although it is important to bear these factors in mind, for ease of direct comparison with the model from \citetalias{Narayanan_2014}, we proceed by considering that the star-formation rate surface densities that we measure are a sufficient proxy for the volume densities, though we discuss some of the physical properties of the galaxies that might correlate with the CO SLED in the next section.

We computed the maximal information coefficient (MIC) to carry out a maximal information-based nonparametric exploration of the other observable properties of the sample for which we have existing information, that may correlate with the offsets we find between the observations and the \citetalias{Narayanan_2014} model predictions of the line ratios. For mid-$J_{\rm up}$ to CO(1--0) line ratios, only one source (AS2UDS012.0) differed from the model by more than $1\sigma$ which did not suffice the use of a statistical test, however, for high-$J_{\rm up}/$mid-$J_{\rm up}$ line ratios, eight sources differed by more than $1\sigma$ and one (AS2COS0065.1) by more than $3\sigma$.
For the latter line ratios, the properties we investigated include the dust temperature ($T_{\rm d}$), dust continuum flux ($S_{870}$), dust luminosity ($L_{\rm d}$), dust mass ($M_{\rm d}$), dust continuum size ($R_{\rm d}$), stellar mass ($M_\ast$), and dust extinction ($A_{\rm V}$). The strongest correlations we find are between the offsets and stellar mass and $S_{870}$, which have ${\rm MIC} \sim 0.7$ and ${\rm MIC} \sim 0.4$, respectively.
While these are noticeable correlations, Jackknife resampling of the data shows that there is at least an $\approx 8\%$ chance of finding a correlation of this strength by chance in our sample, i.e. the strength of the correlation is only significant to $\approx 2\sigma$.






\section{Discussion}
\label{sec:discussion}


There is a disagreement about which is the primary driver of the low-$J_{\rm up}$ excitation in low-redshift, star-forming galaxies.
Using the $R_{5,2}$ line ratio as a proxy for the excitation, \citet{Daddi_2015} showed that the primary driver is the average intensity of the interstellar radiation field, while \citet{Valentino_2020} on the other hand recently showed that it is the star-formation efficiency that correlates more strongly with excitation.
They found that $R_{4,1}$ does not correlate well with star-formation efficiency but correlates with star-formation rate surface density, though they did not investigate the average intensity of the interstellar radiation field. They also suggested that there was a tentative agreement between their results and the predictions from the \citetalias{Narayanan_2014} model.
The low-$J_{\rm up}$ excitation of some of the sources in our sample has already been presented in \citet{FriasCastillo_2023}.

Our results suggest that the high-$J_{\rm up}$ excitation, as traced by the $R_{6,3}$ line ratio, correlates most significantly with the star-formation rate surface density when combining them with observations of low-redshift main-sequence and starburst galaxies, which allows us to test the relationship across nearly three orders of magnitude.
From Figures~\ref{fig:sfe_isrf_models}~and~\ref{fig:model}, it is clear that the low- and high-redshift galaxies overlap in star-formation efficiency and average intensity of the interstellar radiation field, but are separated in star-formation rate surface density.
It seems that the consideration of the compactness of individual galaxies is a critical dimension to distinguish the degrees of excitation exhibited amongst different populations \citep{Puglisi_2021}.
However, while star-formation rate surface density generally correlates with CO excitation, we observe significant scatter in $R_{6,3}$ line ratios at fixed star-formation rate surface density of a factor of $\approx 3$. The constraints we are able to place on the measurements strongly imply that the scatter reflects intrinsic variations in the ISM properties.

The model proposed by \citetalias{Narayanan_2014} accounts for several processes that influence the molecular gas temperature and, by extension, level populations that are exhibited in the diversity of CO SLED shapes found from observations. One such effect is that of increased line optical depth which increases molecular gas temperatures through line trapping. High-$J_{\rm up}$ lines are predicted by their model to have increased optical depths in galaxies with higher star-formation rate surface density, while low-$J_{\rm up}$ lines are predicted to have lower optical depths at higher star-formation rate surface densities.
It is worth noting that different assumptions on how (supersonic) turbulence within giant molecular clouds is modelled may affect the CO line optical depths, and since the optical depth varies with CO transition number, different models for turbulence might yield different line ratios \citep{Bournaud_2015,Kamenetzky_2018,Bisbas_2021}.
Furthermore, the model considers the influence of increased cosmic ray heating rates, which have been proposed to be the main cause of molecular gas heating in starburst nuclei \citep{Suchkov_1993}. Ions and excited molecules produced by cosmic rays can significantly heat the gas and cause temperature gradients in prestellar cores \citep{Glassgold_1973,Dalgarno_1999,Padovani_2009,Glassgold_2012}. \citetalias{Narayanan_2014} employ these effects by assuming that the cosmic ray ionization rate scales linearly with the galaxy-wide SFR \citep{Acciari_2009,Abdo_2010,Hailey_2008a}, finding that its effect diminishes in molecular clouds with very high gas densities ($n \gg 10^4$\,cm$^{-3}$) and galaxies with higher $\Sigma_{\rm SFR}$.
Another source of molecular gas heating which is neglected by the model is turbulence dissipation via low-velocity shocks, expected to be of the same order as cosmic ray heating and to significantly affect the excitation conditions of the ISM \citep{Rosenberg_2014,Rosenberg_2015,Harrington_2021}.
Alternative models including turbulence heating were used to investigate their effects on molecular clouds via both adiabatic compression and viscous dissipation in \citet{Narayanan_2011}, finding only modest differences to the gas temperatures and, as a consequence, the CO SLEDs. However, different prescriptions to model the heating rate expected from turbulence dissipation for unresolved GMCs in their simulations change the predicted properties of the molecular gas.
\citet{Pon_2012} found that these shocks can heat molecular gas above 100~K which can cause higher rotational states to become more populated and, despite heating only 1\% of the gas, can contribute more emission to the $J_{\rm up} = 5$ and higher lines than unshocked gas. Photon-dominated region (PDR) models have also been used to show that low- and high-$J_{\rm up}$ CO lines are sensitive to the mechanical heating effect of turbulence dissipation by differing amounts, which may cause strong changes in line ratios \citep{Kazandjian_2012,Kazandjian_2015}.
It is possible that an assumption of the relationship between cosmic ray ionization rates and galaxy-wide star-formation rate, in combination with the neglected heating effect by turbulence dissipation, is enough to cause the scatter around the model which is seen in the data, especially towards high star-formation rate surface densities.



It is likely that the CO SLED of SMGs is multi-variable in nature, where the power sources heating the ISM probably consist of photons, cosmic-rays, and turbulence. For the same star-formation rate density, these can play dramatically different roles depending on how compact the star-forming region is, and whether the gas dynamics are dominated by rotation or a merger.

Particularly, the contribution of cosmic rays to ISM heating depends largely on their acceleration, driven by processes like supernova shocks and AGN jets. These determine whether cosmic rays primarily lose energy via hadronic interactions, producing gamma rays that escape the system, or through the streaming instability mechanism, where cosmic rays scatter off magnetic field irregularities and transfer energy thermally to the ISM through Alfv\'{e}n waves.
However, the gamma-ray spectra of starburst galaxies such as Arp 220, NGC 253, and M82 suggest that the streaming effect is not dominant, as these systems exhibit hard GeV spectra that soften at higher energies, indicating that cosmic rays primarily lose energy through hadronic collisions with gas rather than streaming \citep{Krumholz_2023}. This trend implies that cosmic rays may not be able to significantly heat the ISM, at least in low-redshift starburst galaxies.
However, in the densest and most highly star-forming regions, at high redshift, the cosmic rays may play a more important role \citep{Papadopoulos_2011}.

Furthermore, the effect of turbulence on measured CO line ratios may be significant in the context of starburst galaxies \citep{Papadopoulos_2012}, where high star-formation rates can generate turbulence through supernova explosions and gravitational instabilities.
High turbulence can lead to enhanced collisional excitation, which can increase the intensity of high-$J$ lines relative to mid-$J$ lines, thereby altering the line ratios observed.
Turbulence is also expected to play a crucial role in altering low-$J$ line ratios, even though it may be secondary to the more immediate and substantial impacts of the interstellar radiation field and cosmic ray ionization rate \citep{Penaloza_2017}.

Since the optical depths of different CO transitions in the \citetalias{Narayanan_2014} model vary significantly with respect to star-formation rate surface density, obtaining multi-$J_{\rm up}$ $^{13}$CO observations would help to constrain them and allow the model to better reproduce the observed CO SLEDs.

Finally, we note that two-component gas density models have been shown to be a good description of CO SLEDs in $z = $\,1--3 DSFGs \citep{Harrington_2021}.
In this scenario, high-$J_{\rm up}$ excitation is driven by higher gas densities than the lower-$J_{\rm up}$ excitation, and gas kinetic temperatures are likely higher in component two tracing the denser, higher-$J_{\rm up}$ excited gas.
Since we measure high $r_{3,1}$ and low $r_{6,3}$ line ratios, a broad two-phase differentiation of the molecular gas is certainly apparent in SMGs.

\section{Conclusions}
\label{sec:conclusions}

We analysed new ALMA observations of the CO(6--5) or CO(8--7) line in combination with CO(3--2) or CO(4--3) and JVLA observations of CO(1--0) for twelve $z =$\,2--4 SMGs. By exploiting existing measurements of high-resolution (0.8 arcsec) $870$\,$\upmu$m dust sizes and star-formation rates, as well as dust mass measurements determined from multi-wavelength SED fitting, we investigated the parameterisation of the CO SLED to test the relationship between SLED shape and physical drivers, with a particular focus on star-formation rate surface density ($\Sigma_{\rm SFR}$). Our sample alone has a range in $\Sigma_{\rm SFR}$ spanning an order of magnitude.

Our conclusions can be summarised as follows:
\begin{enumerate}
    \item The CO(6--5) line luminosities are consistent with other $z > 1$ SMGs at similar far-infrared luminosities to within $2\sigma$, and the $L_{\rm IR}$ vs $L^{\prime}_{\rm CO(6-5)}$ relation agrees with local ULIRGs, Seyfert/spiral nuclei, nearby galaxies, and $z > 1$ SMGs.
    \item The typical SLED of a dusty, luminous SMG normalised to the CO(1--0) is more excited at $J_{\rm up} = 3$ than the Milky Way inner disc by a factor of $\approx 3$, and is akin to the SLEDs of the nearby starburst galaxies NGC 253 and M82, and local ULIRGs.
    There is a large scatter of a factor of $\approx 3$ in the line ratios at $J_{\rm up} = 6$, which allows us to test the relationship between high-$J_{\rm up}$ CO excitation and physical drivers.
    \item 
    Our sample extends the ranges of star-formation rate surface density from previous investigations of the physical drivers of CO excitation in low-redshift, star-forming galaxies to cover three orders of magnitude, while sharing similar ranges in star-formation efficiency and average intensity of the interstellar radiation field.
    On average, the $R_{6,3}$ line ratios, used as a proxy for excitation, are a factor of $\approx 2.5$ times greater than lower luminosity samples, suggesting that the ISMs of SMGs may consist of more intense conditions on average than low-redshift, less actively star-forming galaxies.
    The elevated excitation levels of our observations compared to the extrapolated fit to the observations in \citet{Valentino_2020}, suggests that the $R_{6,3}$--$\Sigma_{\rm SFR}$ relationship may be different in low-redshift, less active systems and high-redshift, highly star-forming systems.
    \item The strongest driver of high-$J_{\rm up}$ CO excitation, in star-forming galaxies, when also considering low-redshift main-sequence and starburst galaxies, is $\Sigma_{\rm SFR}$.
    The compactness of individual galaxies seems to be a critical dimension to distinguish the differences in excitation exhibited amongst different populations, since it separates the low-redshift main-sequence and starburst galaxies from high-redshift SMGs.
    \item The CO SLED shape in SMGs is consistent with the \citetalias{Narayanan_2014} unresolved empirical model (with added extinction) to within $3\sigma$. However we find significant scatter of a factor of $\approx 3$ in the $R_{6,3}$ line ratios that is not explained solely by systematics and may represent intrinsic variations in the ISM of SMGs, such as cosmic ray ionization rates and mechanical heating effects through turbulence dissipation.
    Spatially resolved observations, and their comparison to the \citetalias{Narayanan_2014} model are needed to test this.
    \item We investigate how other observable properties of the sample correlate with the offsets between the observations and model predictions of the high-$J_{\rm up}/$mid-$J_{\rm up}$ line ratios. We find tentative correlations between the offsets and stellar mass and $S_{870}$.
    \item We test the form of the relationship between $R_{6,3}$ and $\Sigma_{\rm SFR}$ by comparing a linear fit and power-law fit to the quantities on logarithmic scales, finding that a linear fit is preferred.
\end{enumerate}

Studies of the dependence of the shape of CO SLEDs so far have been limited in terms of the precision and range of measured star-formation rate surface densities, to low-$J_{\rm up}$ CO transitions and low redshifts. This study has made use of the impressive capabilities of ALMA to improve upon each of these. 
Future theoretical CO SLED models would greatly benefit from multi-$J_{\rm up}$ $^{13}$CO observations to constrain the average line optical depths that aid in classifying CO SLEDs at $J_{\rm up} \geq 3$.
Higher spatial resolution observations of CO lines, including $J_{\rm up} \geq 7$ for identifying the peak of the SLEDs, will be essential in determining if global variations in the CO excitation of star-forming galaxies highlighted in this work can be attributed to variations on the local scales of giant molecular clouds.

\section*{Acknowledgements}

The authors thank the anonymous referee for their helpful and insightful comments which have greatly improved the paper. D.J.T acknowledges the support of Science and Technology Facilities Council (STFC) studentship (ST/X508354/1). A.M.S, I.R.S, and Z.L acknowledge STFC consolidated grant ST/X001075/1.
This paper makes use of the following ALMA data: ADS/JAO.ALMA\#2016.1.00564.S, \#2017.1.01163.S, \#2017.1.01512.S, \#2019.1.00337.S, \#2019.1.01600.S, and \#2021.1.00666.S. ALMA is a partnership of ESO (representing its member states), NSF (USA) and NINS (Japan), together with NRC (Canada), NSTC and ASIAA (Taiwan), and KASI (Republic of Korea), in cooperation with the Republic of Chile. The Joint ALMA Observatory is operated by ESO, AUI/NRAO and NAOJ. The National Radio Astronomy Observatory is a facility of the National Science Foundation operated under cooperative agreement by Associated Universities, Inc. C.-C.C. acknowledges support from the National Science and Technology Council of Taiwan (111-2112M-001-045-MY3), as well as Academia Sinica through the Career Development Award (AS-CDA-112-M02). M. R. is supported by the NWO Veni project ”Under the lens” (VI.Veni.202.225). M.F.C. acknowledges support of the VIDI research programme with project number 639.042.611, which is (partly) financed by the Netherlands Organisation for Scientific Research (NWO). E.F.-J.A. acknowledge support from UNAM-PAPIIT project IA102023, and from CONAHCyT Ciencia de Frontera project ID: CF-2023-I-506. 


\section*{Data Availability}

The data used in this work was taken using the Atacama Large Millimeter-submillimeter Array (ALMA), accessible through the ALMA science archive at \url{https://almascience.eso.org}, and the Karl G. Jansky Very Large Array (JVLA), accessible through the National Radio Astronomy Observatory data archive at \url{https://data.nrao.edu/portal/#/}.



\bibliographystyle{mnras}
\bibliography{paper} 




\appendix
\setcounter{table}{0}
\renewcommand{\thetable}{A\arabic{table}}
\setcounter{figure}{0}
\renewcommand{\thefigure}{A\arabic{figure}}

\begin{table*}
\centering
    \caption{Target ALMA ID, CO line of the emission, observed frequency of transition, synthesised beam FWHM (major-axis diameter) of the data-cube for the transition analysed in this work, signal-to-noise ratio of the fitted spectrum, dust emissivity spectral index of the continuum spectrum, FWHM of the Gaussian profile fit to the line spectra, measured flux from the fit to the spectra, and corresponding line luminosity.}
    \begin{tabular}{@{}ccccccccc@{}}
		\hline\hline
		Target & $^{12}$CO transition & $\nu_{\rm obs}$ & $\theta_{\rm maj} \times \theta_{\rm min}$ & S / N & $\beta$ & FWHM$^e$ & ${I_{\rm CO}}^f$ & ${L^{\prime}_{\rm CO}}^f$ \\
         &  & (GHz) & (arcsec) & & & (km s$^{-1}$) & (Jy km s$^{-1}$) & ($10^{10}$ K km s$^{-1}$ pc$^2$) \\
		\hline
        AS2UDS009.0  & $J = 1\rightarrow 0$ & 29.245  & -                  & 2  & -    & $480 \pm 230$                & $0.12 \pm 0.08$      & $4.81 \pm 3.20$\\
                     & $J = 3\rightarrow 2$ & 87.730  & -                  & 5  & -    & $460 \pm 90$, $270 \pm 50$   & $1.17 \pm 0.24^g$ & $5.23 \pm 1.08$\\
                     & $J = 6\rightarrow 5$ & 175.430 & $0.72 \times 0.52$ & 36 & 1.36 & $460 \pm 110$, $270 \pm 110$ & $1.89 \pm 0.17^g$ & $2.10 \pm 0.19$\\
        \hline
        AS2UDS011.0  & $J = 1\rightarrow 0$ & 22.724  & -                  & 2  & -   & $840 \pm 340$  & $0.07 \pm 0.04$ & $4.75 \pm 2.71$\\
                     & $J = 4\rightarrow 3$ & 90.889  & -                  & 6  & -   & $690 \pm 120$  & $0.81 \pm 0.14$ & $3.43 \pm 0.59$\\
                     & $J = 6\rightarrow 5$ & 136.316 & $1.49 \times 0.98$ & 25 & 1.5 & $690 \pm 90$   & $1.43 \pm 0.17$ & $2.71 \pm 0.32$\\
        \hline
        AS2UDS012.0  & $J = 1\rightarrow 0$ & 32.747  & -                  & 5  & -   & $860 \pm 200$  & $0.36 \pm 0.07$ & $11.1 \pm 2.16$\\
                     & $J = 3\rightarrow 2$ & 98.237  & -                  & 5  & -   & $620 \pm 120$  & $1.12 \pm 0.21$ & $3.82 \pm 0.71$\\
                     & $J = 6\rightarrow 5$ & 196.440 & $1.88 \times 1.29$ & 21 & 2.2 & $620 \pm 50$   & $1.78 \pm 0.21$ & $1.52 \pm 0.18$\\
        \hline
        AS2UDS014.0  & $J = 1\rightarrow 0$ & 23.997  & -                  & -  & -   & -             & $< 0.12$        & $< 7.31$       \\
                     & $J = 4\rightarrow 3$ & 95.980  & -                  & 7  & -   & $910 \pm 130$ & $1.85 \pm 0.25$ & $7.04 \pm 0.96$\\
                     & $J = 6\rightarrow 5$ & 143.951 & $0.95 \times 0.82$ & 33 & 1.3 & $910 \pm 80$  & $1.59 \pm 0.22$ & $2.69 \pm 0.37$\\
        \hline
        AS2UDS026.0  & $J = 1\rightarrow 0$ & 26.834  & -                  & 2  & -   & $200 \pm 170$ & $0.13 \pm 0.05$ & $6.28 \pm 2.42$\\
                     & $J = 4\rightarrow 3$ & 107.327 & -                  & 7  & -   & $690 \pm 100$ & $1.34 \pm 0.20$ & $4.05 \pm 0.61$\\
                     & $J = 6\rightarrow 5$ & 160.970 & $1.14 \times 0.96$ & 16 & 1.4 & $690 \pm 50$  & $1.24 \pm 0.26$ & $1.66 \pm 0.35$\\
        \hline
        AS2UDS072.0  & $J = 1\rightarrow 0$ & -  & -                  & -  & -   & -             & -        & -        \\
                     & $J = 4\rightarrow 3$ & 101.525 & -                  & 7  & -   & $320 \pm 50$  & $1.44 \pm 0.26$ & $4.90 \pm 0.87$\\
                     & $J = 8\rightarrow 7$ & 203.015 & $0.79 \times 0.66$ & 15 & 2.7 & $320 \pm 30$  & $2.06 \pm 0.26$ & $1.75 \pm 0.22$\\
        \hline
        AS2UDS126.0  & $J = 1\rightarrow 0$ & 33.545  & -                  & 4  & -      & $460 \pm 120$ & $0.17 \pm 0.10$ & $4.94 \pm 2.91$\\
                     & $J = 3\rightarrow 2$ & 100.630 & -                  & 8  & -      & $630 \pm 80$  & $1.53 \pm 0.20$ & $4.95 \pm 0.65$\\
                     & $J = 6\rightarrow 5$ & 201.225 & $0.74 \times 0.59$ & 17 & $-0.4$ & $630 \pm 120$ & $2.93 \pm 0.53$ & $2.37 \pm 0.43$\\
        \hline
        AS2COS0009.1 & $J = 1\rightarrow 0$ & 35.360  & -                  & 0.2  & -   & -                          & $< 0.14$             & $< 3.57$       \\
                     & $J = 3\rightarrow 2$ & 106.076 & -                  & 17   & -   & $320 \pm 20$, $220 \pm 10$   & $1.54 \pm 0.09^g$ & $4.38 \pm 0.26$\\
                     & $J = 6\rightarrow 5$ & 212.115 & $0.89 \times 0.70$ & 25   & 4.6 & $320 \pm 230$, $220 \pm 170$ & $2.57 \pm 0.49^g$ & $1.82 \pm 0.35$\\
        \hline
        AS2COS0014.1 & $J = 1\rightarrow 0$ & 29.397  & -                  & 6  & -   & $570 \pm 100$              & $0.23 \pm 0.09$      & $9.11 \pm 3.56$\\
                     & $J = 3\rightarrow 2$ & 88.186  & -                  & 7  & -   & $310 \pm 40$, $380 \pm 50$ & $1.59 \pm 0.23^g$ & $6.98 \pm 1.02$\\
                     & $J = 6\rightarrow 5$ & 176.342 & $1.08 \times 0.73$ & 47 & 4.6 & $310 \pm 30$, $380 \pm 40$ & $1.79 \pm 0.11^g$ & $1.97 \pm 0.12$\\
        \hline
        AS2COS0044.1 & $J = 1\rightarrow 0$ & 32.205  & -                  & 0.2  & -   & -                          & $< 0.17$             & $< 5.45$       \\
                     & $J = 3\rightarrow 2$ & 96.610  & -                  & 13   & -   & $420 \pm 30$, $350 \pm 30$   & $0.77 \pm 0.06^g$ & $2.76 \pm 0.21$\\
                     & $J = 6\rightarrow 5$ & 193.187 & $0.73 \times 0.62$ & 61   & 1.4 & $420 \pm 240$, $350 \pm 260$ & $0.82 \pm 0.27^g$ & $0.73 \pm 0.24$\\
        \hline
        AS2COS0065.1 & $J = 1\rightarrow 0$ & 33.764  & -                  & 2  & -      & $290 \pm 190$              & $0.10 \pm 0.15$      & $2.86 \pm 4.29$\\
                     & $J = 3\rightarrow 2$ & 101.288 & -                  & 20 & -      & $530 \pm 30$, $240 \pm 10$ & $1.35 \pm 0.07^g$ & $4.28 \pm 0.21$\\
                     & $J = 6\rightarrow 5$ & 202.540 & $0.74 \times 0.60$ & 35 & $-0.5$ & $530 \pm 90$, $240 \pm 50$ & $3.66 \pm 0.23^g$ & $2.91 \pm 0.18$\\
        \hline
        AS2COS0139.1 & $J = 1\rightarrow 0$ & 26.855  & -                  & 4  & -      & $440 \pm 100$ & $0.18 \pm 0.09$      & $8.68 \pm 4.34$\\
                     & $J = 4\rightarrow 3$ & 107.411 & -                  & 18 & -      & $480 \pm 30$  & $1.46 \pm 0.11^g$ & $4.42 \pm 0.32$\\
                     & $J = 8\rightarrow 7$ & 214.757 & $0.93 \times 0.74$ & 27 & $-0.3$ & $480 \pm 50$  & $2.86 \pm 0.40^g$ & $2.16 \pm 0.30$\\
		\hline\hline
	\end{tabular}
    \captionsetup{labelformat=notes}
    \caption*{
    
    $^e$Mid-$J_{\rm up}$ FWHM errors computed using the S/N of the spectra. High-$J_{\rm up}$ FWHM errors determined using a Monte Carlo simulation. The FWHM of each Gaussian component is separated by a comma.
    
    $^f$Existing CO(1--0) measurements or $2\sigma$ lower limits have been taken from \citet{FriasCastillo_2023} and Jansen et al. (2024, in prep.), while CO(3--2) and CO(4--3) observations come from \citet{Birkin_2021} and \citet{Chen_2022} for AS2UDS and AS2COSMOS sources, respectively.
    
    $^g$Measured from the sum of a double Gaussian fit to the lines. 
    
    We note that for sources AS2UDS072.0 and AS2COS0139.1, the emission line in our study is taken to be CO(8--7) rather than CO(6--5), due to a misidentification of the mid-$J_{\rm up}$ CO line.}
    \captionsetup{labelformat=default}
    \label{tab:fluxes}
\end{table*}

\begin{figure}
\centering
	\includegraphics[width=1\columnwidth]{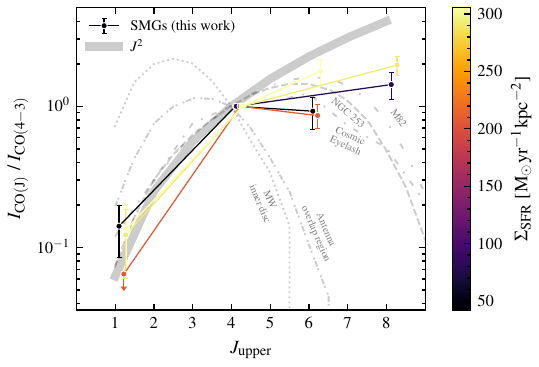}
    \caption{CO SLEDs for five of our sample SMGs with available CO(4--3) fluxes, coloured by their respective star-formation rate surface densities denoted by the colour bar. Individual fluxes measured from fits to the spectra are shown as circles with their error-bars indicating their propagated uncertainties determined via a Monte Carlo method, and are normalised to the CO(4--3) line. 
    We add a small offset to the $J_{\rm up}$ of each measurement to separate their errorbars which would otherwise overlap.
    The solid-grey line shows $J_{\rm up}^2$ scaling, the expected scaling of intensities for levels in LTE and in the Rayleigh-Jeans limit. The CO SLEDs of other known systems are shown in the background as different grey-dashed lines. For the sources shown, none are considered by us to be potential AGN hosts.}
    \label{fig:extra_sled}
\end{figure}

\bsp	
\label{lastpage}
\end{document}